\documentclass[lettersize,journal]{IEEEtran}
\IEEEoverridecommandlockouts

\usepackage{mathrsfs}
\usepackage[noadjust]{cite}
\usepackage{graphicx,color,overpic,psfrag}
\usepackage{amsmath, amssymb}
\usepackage{algorithm}  
\usepackage{algpseudocode} 
\usepackage{latexsym}
\usepackage{bm}
\usepackage{cases}
\usepackage{array}
\usepackage{fancyhdr}
\usepackage{setspace}
\usepackage{float}
\usepackage{caption}
\UseRawInputEncoding
\usepackage{indentfirst}
\usepackage{url}
\usepackage{subfigure}  
\usepackage{blkarray}
\usepackage{booktabs}
\usepackage{multirow}
\usepackage{dsfont}
\usepackage{tabularx}
\usepackage[table]{xcolor}
\usepackage{amsfonts}
\usepackage{amsthm}
\usepackage{stfloats}
\usepackage{letltxmacro}
\usepackage{epsfig}
\usepackage{epstopdf} 
\usepackage {color}
\usepackage{multicol}

\newcommand{\ra}[1]{\renewcommand{\arraystretch}{#1}}

\newcommand{\figref}[1]{Fig. \ref{#1}}
\newcommand{\tabref}[1]{Table \ref{#1}}

\newcommand{\secref}[1]{Section \ref{#1}}

\newcommand{\fmin}{f_{1}}
\newcommand{\fmax}{f_{2}}

\newcommand{\frei}{f_{i}}

\newcommand{\steer}{\mathbf{a}}

\newcommand{\rr}{\mathbf{R}}
\newcommand{\mean}{\mathbb{E}}

\graphicspath{{figure/}}
\allowdisplaybreaks

\newcommand{\pn}{\Psi}
\newcommand{\gis}{\bm{\mathcal{G}}_{i}(S)}
\newcommand{\gjs}{\bm{\mathcal{G}}_{j}(S)}
\begin{document}
	
	\title{CF-CGN: Channel Fingerprints Extrapolation for Multi-band Massive MIMO Transmission based on Cycle-Consistent Generative Networks}

	\author{  Chenjie~Xie,~\IEEEmembership{Graduate Student~Member,~IEEE,}
		           Li~You,~\IEEEmembership{Senior~Member,~IEEE,}\\
		          Zhenzhou~Jin,~\IEEEmembership{Graduate Student~Member,~IEEE,}
		          Jinke~Tang,~\IEEEmembership{Graduate Student~Member,~IEEE,}
		          Xiqi~Gao,~\IEEEmembership{Fellow,~IEEE,} and
		          Xiang-Gen~Xia,~\IEEEmembership{Fellow,~IEEE}

		\thanks{An earlier version of this paper was submitted to IEEE INFOCOM \cite{00}.
			
			Chenjie Xie, Li You, Zhenzhou Jin, Jinke Tang, and Xiqi Gao are with the National Mobile Communications Research Laboratory, Southeast University, Nanjing 210096, China, and also with the Purple Mountain Laboratories, Nanjing 211111, China (e-mail: cjxie@seu.edu.cn, lyou@seu.edu.cn, zzjin@seu.edu.cn, jktang@seu.edu.cn, xqgao@seu.edu.cn).
			
			Xiang-Gen Xia is with the Department of Electrical and Computer Engineering, University of Delaware, Newark, DE 19716, USA (e-mail: xianggen@udel.edu).		 
		}
	}

	\maketitle
	\thispagestyle{empty}
\begin{abstract}
Multi-band massive multiple-input multiple-output (MIMO) communication can promote the cooperation of licensed and unlicensed spectra, effectively enhancing spectrum efficiency for Wi-Fi and other wireless systems.
As an enabler for multi-band transmission, channel fingerprints (CF), also known as the channel knowledge map or radio environment map, are used to assist channel state information (CSI) acquisition and reduce computational complexity.
In this paper, we propose CF-CGN (Channel Fingerprints with Cycle-consistent Generative Networks) to extrapolate CF for multi-band massive MIMO transmission where licensed and unlicensed spectra cooperate to provide ubiquitous connectivity.
Specifically, we first model CF as a multichannel image and transform the extrapolation problem into an image translation task, which converts CF from one frequency to another by exploring the shared characteristics of statistical CSI in the beam domain. Then, paired generative networks are designed and coupled by variable-weight cycle consistency losses to fit the reciprocal relationship at different bands. Matched with the coupled networks, a joint training strategy is developed accordingly, supporting synchronous optimization of all trainable parameters. During the inference process, we also introduce a refining scheme to improve the extrapolation accuracy based on the resolution of CF.
Numerical results illustrate that our proposed CF-CGN can achieve bidirectional extrapolation with an error of 5 $\sim$ 17 dB lower than the benchmarks in different communication scenarios, demonstrating its excellent generalization ability. We further show that the sum rate performance assisted by CF-CGN-based CF is close to that with perfect CSI for multi-band massive MIMO transmission.
\end{abstract}
	
	
\section{Introduction}
    With the continuous growth of demands for wireless communications, the sixth generation (6G) communication systems have attracted worldwide attention, being expected to provide high-quality communications in the future \cite{01}. However, the scarcity of spectrum resources has become increasingly prominent, posing bottlenecks for the development of wireless communications \cite{02}. In the foreseeable future, technologies operating in both licensed and unlicensed spectra, such as multi-band massive multiple-input multiple-output (MIMO), will play a pivotal role in alleviating this issue and supporting ubiquitous connectivity \cite{03}.

	
    For Wi-Fi and other wireless systems, multi-band massive MIMO serves as an enabler to integrate different spectra, showing significant potential to enhance spectrum efficiency \cite{02, 03}. Meanwhile, combining various bands enables ultra-high-speed transmission with ubiquitous coverage due to the strong penetration capability of low frequencies and the significant bandwidth of high frequencies, better fulfilling the needs of different scenarios \cite{08}. However, several issues still need to be solved regarding the recent multi-band massive MIMO transmission. For instance, in order to acquire the statistical channel state information (CSI), base stations (BS) or access point (AP) need to consider the coordination among different frequencies and perform channel estimation across all of them, which is more challenging and will increase significant computation burden \cite{06}, \cite{07}. 
	
\begin{figure}[htbp]
		\centerline
		{\includegraphics[width=1\linewidth]{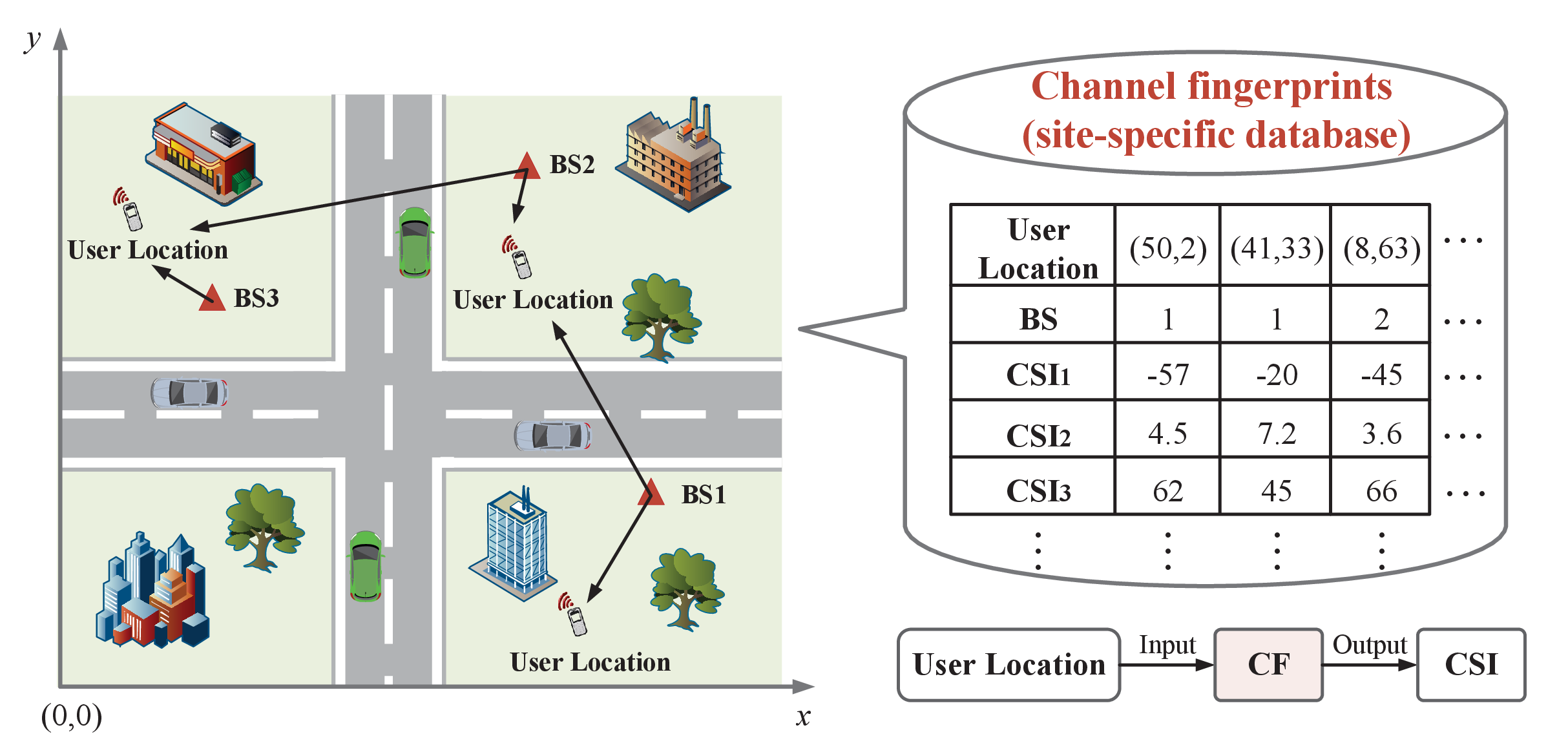}	}	
		\caption{Schematic diagram of channel fingerprints.}
		\label{CFs}
	\end{figure}

	Fortunately, channel fingerprints (CF), also known as the radio environment map (REM) \cite{rem} or the channel knowledge map (CKM) \cite{ckm}, have been proposed as a new solution to the above issues, which can be regarded as a site-specific database containing the user-location-related channel information. The schematic diagram is shown in \figref{CFs}. Due to the slow-changing characteristic of the statistical CSI, we can store it as the target channel information in CF and thus assist in beam selection, power assignment, Wi-Fi location, and other applications without repetitive estimations. In multi-band massive MIMO systems, where channels of different bands share similarities due to the same electromagnetic propagation environment \cite{re18}, the application of CF enables BS to flexibly schedule licensed and unlicensed spectra on demand, providing new development insights for 5G NR-U, particularly in the context of Wi-Fi technology.

	Recently, numerous studies have been conducted on channel fingerprints construction. Based on the cylindrical parabolic equation model (CPEM), the distribution of local radiation was first calculated in \cite{rem1}, and the results were then transferred to global fine grids (FG) for REM estimation in a large-scale region. The Kriging-based method is leveraged in \cite{kriging} and \cite{kg2} to track channel gains in a specified geographical region, which is denoted as the channel gain map (CGM). To break through the spatial resolution limitations, authors in \cite{kg3} enhanced the prior observations by interpolation and then introduced two uncertainty-aware matrix completion algorithms to reconstruct the propagation map. Apart from these traditional interpolation methods, the Gaussian process (GP) is also adopted to construct the REM. Specifically, authors in \cite{gp1} considered the positional uncertainty and developed an algorithm based on variational inference, and authors in \cite{gp2} introduced K-nearest neighbors (KNN) in GP to increase the accuracy of REM construction in urban scenarios.
	
	As the data dimension expands, algorithmic overheads required for these traditional methods experience exponential growth, so researchers have tried to adopt machine learning (ML) \cite{ml1, ml3, ml2, ml4} to solve the problem. With the aid of the well-established channel models, the author reconstructed CGM based on the expectation maximization (EM) algorithm \cite{em}. Authors in \cite{gnn} transformed the spatially sparse measurements into a graph and utilized the graph neural network (GNN) to construct the radio map. Focusing on the scalability and privacy limitations, authors in \cite{lb} introduced a long short-term memory (LSTM) model based on federated learning to estimate the indoor coverage map. Laplacian pyramid (LP) was also applied to reconstruct CKM, which comprises several sub-networks for different frequency components \cite{LP, lpp}. In \cite{aun}, the authors estimated the global path loss using convolutional neural networks (CNNs) and named it RadioUNet. Apart from these classical networks, generative models such as auto-encoders (AE) and generative adversarial networks (GANs) have also been explored for application. For example, authors in \cite{da} developed a UNet-based deep completion AE to generate REM of shadowing, which maintains good performance even with a smaller number of measurements. Authors in \cite{gan1} combined GAN with ResNet to recover the received signal strength (RSS), and authors in \cite{gan2} designed a conditional GAN with a two-phase learning framework to estimate the CGM.                  
	
    Although these generative AI-based methods have presented excellent performance in terms of the CF construction, most of them are limited to single-band systems, which do not apply to multi-band transmission because of the neglect of coordination between different bands. In practice, it has been confirmed that there exists a specific correlation among different bands in multi-band systems due to the consistence of the physical environment and channel propagation characteristics \cite{xgx, xgx3}, so the prior information in one band can be used to construct CF in another band. 
    Based on this idea, techniques like spectrum cartography attempt to construct CF for multi-band systems. For example, a dynamic compressed sensing (DCS) based framework was proposed in \cite{df1} to estimate the power spectral density (PSD) map. Authors in \cite{df2} and \cite{df3} presented a constrained block-term tensor decomposition (CBTD) method to jointly recover the PSD and spatial loss field (SLF). However, these strategies ignore the integration of licensed and unlicensed spectra and still focus on CF with a certain kind of CSI, typically represented by path loss or signal strength, which may not be able to well characterize the features of channels further. Another feasible method is to utilize the autoregressive (AR) model to establish the mapping of power angle spectrum (PAS) between different frequencies for individual users in \cite{tjk} and its longer version \cite{tjkk}. However, if we apply this approach to the extrapolation of CF, it needs to traverse the location of each potential user, resulting in an unacceptable computational burden.
	
    Motivated by the above discussions, we investigate the generative models-based CF extrapolation method for multi-band massive MIMO transmission where licensed and unlicensed spectra cooperate to provide ubiquitous connectivity. Note that the CF we construct here is not necessarily limited to only a single channel information but rather an aggregation of numerous statistical CSI, such as RSS, coverage, line of sight (LOS), or even the channel covariance matrix. Since CF can be considered a multichannel image, the extrapolation problem is formulated as the image translation task in computer vision (CV), which aims to convert CF from one frequency to another based on the shared characteristics of statistical CSI in the beam domain. Given the excellent performance and widespread application of generative models in both CF construction problems and image translation tasks, the paired generative networks are designed, which extract features from CF at one frequency, map the data manifolds into a latent space, and then extrapolate CF to another frequency through the underlying correlations. Additionally, it is essential but challenging to establish a reciprocal relationship among CF in different bands so that they can be predicted mutually, thus ensuring a one-to-one correspondence between any paired frequencies. Researchers in the CV field have made some efforts and put forward the concept of cycle consistency, which aims to achieve the mutual transformation of images between two domains while retaining their original features \cite{cg2, cg3}. Based on this idea, we couple the networks through a variable-weight cycle consistency loss to fit the reciprocal relationship and develop a joint training strategy to optimize the trainable parameters synchronously. The main contributions of this paper can be summarized as follows:
	\begin{itemize}
	 	\item We investigate the CF model based on the channel model that reveals a relationship between the spatial covariance matrix and channel PAS. By modeling CF as a multichannel image, we formulate the extrapolation problem as an image translation task, which converts CF from one frequency to another while retaining its inherent characteristics.
    	\item Based on the shared features of CF at different bands, we adopt cycle-consistent generative networks to extrapolate CF for multi-band massive MIMO systems, named CF-CGN. In particular, a pair of generative networks is designed and coupled by variable-weight cycle consistency losses to fit the CF's reciprocal relationship at different bands. Then, a joint training strategy is developed accordingly, supporting synchronous optimization of all trainable parameters. During the inference process, we introduce a refining scheme to improve the extrapolation accuracy based on the CF pixel-grid conversion.
	    \item We present numerical results to illustrate that the proposed CF-CGN achieves bidirectional extrapolation with an error of 5 $\sim$ 17 dB lower than the benchmarks and has excellent generalization ability under different communication scenarios. We also conduct experiments to show that the sum rate performance assisted by CF-CGN-based CF is close to that with perfect CSI for multi-band massive MIMO transmission.
	\end{itemize}
    
    The rest of this paper is organized as follows. In Section \ref{mode}, we establish the channel model and CF model for multi-band massive MIMO systems and formulate the problem accordingly. Then, in Section \ref{CU}, we introduce the structure of cycle-consistent generative networks and explain the training strategy and refining scheme. Numerical results are presented in Section \ref{nr}. Finally, we conclude the paper in Section \ref{sec_conclusion}.     
    
    \textit{Notations}: $\bar{\jmath} = \sqrt{-1}$ is the imaginary unit. $\mean\{\cdot\}$ denotes the expectation operation, ${\rm{tr}} \{\cdot\}$ represents the trace and $(\cdot)^{*}$ means the conjugate. The notation $\triangleq$ is used for definitions, and $\otimes$ means Kronecker product. Additionally, $\bm{A}^{H}$ represents the conjugate transpose of matrix $\bm{A}$, $\bm{a}^{T}$ represents the transpose of vector $\bm{a}$, $\left[ \bm{a}\right]_{m} $ is the $m$-th element of $\bm{a}$ and $\| A \|_{F}$ is the Frobenius norm. Meanwhile, $\delta(\cdot)$ denotes the Dirac delta function or impulse function, $\circ$ represents the composition of functions, mod is the modulo operation, and $\lceil (\cdot)\rceil$, $\lfloor (\cdot)\rfloor$ represent the ceiling function and floor function, respectively.

	\section{Multi-band System Model and Problem Formulation} \label{mode}
In this section, we introduce the channel model and analyze the relationship between the spatial covariance matrix and PAS in multi-band massive MIMO systems. Based on a specific correlation of PAS at different frequencies, we establish the CF model and formulate the extrapolation problem as an image translation task.
	
	\subsection{Multi-band Channel Model} \label{channelmodel}	
Consider a multi-band massive MIMO system. The BS is equipped with multiple antenna arrays working in $N$ different frequencies, which are denoted as $\mathcal{F}=\{f_{1}, \dots, f_{N}\}$. With respect to the frequency $f_{i} \in \mathcal{F}$, a set of uniform planar arrays (UPA) are arranged at the BS with antenna elements being $M_{i} = M_{i, y} \times M_{i, z}$. We assume that all the arrays that belong to the same BS are co-located and communicate with a plurality of user terminals (UTs) simultaneously, whose potential locations $\bm{q}$ are denoted by the set $\mathcal{Q}$. For simplicity, the UT located in the target area employs a single antenna at each band. In this case, the elevation and azimuth angles of departure are described as $ \theta $ and $ \varphi$, respectively.
    For the BS and the potential UT with location $\bm{q} \in \mathcal{Q}$, the downlink (DL) channel between BS and UT-$ \bm{q} $ at frequency $\frei$ can be modeled as \cite{mod}, \cite{mod2}
	\begin{equation}
	\mathbf{h}_{\bm{q}, i} = \int\limits_{\theta_{\rm min}}^{\theta_{\rm max}} \int\limits_{\varphi_{\rm min}}^{\varphi_{\rm max}}  \beta_{\bm{q}, i} (\theta, \varphi)   \steer_{i}(\theta, \varphi ) d\varphi d\theta \in \mathbb{C}^{M_{i}\times 1},
		\label{channel}
	\end{equation}
	where $ \beta_{\bm{q}, i}(\theta, \varphi ) $ is related to the complex-valued channel gain and $ \steer_{i}(\theta, \varphi ) \in \mathbb{C}^{ M_{i} \times 1} $ is the steering vector. When UPA is deployed, $ \steer_{i}(\theta, \varphi ) = \steer_{z, i}(\theta ) \otimes \steer_{y, i}(\theta, \varphi)$. Define $ \xi_{i} \triangleq d_{i} / \lambda_{i}$ where $ d_{i} $ and $ \lambda_{i} $ are, respectively, the inter-antenna spacing and the wavelength, the $m$-th element of $ \steer_{z, i}(\theta) $ and the $n$-th element of $ \steer_{y, i}(\theta, \varphi) $ can be expressed, respectively, as 
	\begin{align}
		\left[  \steer_{z, i}(\theta)  \right]_{m} &= e^{ - \bar{\jmath} 2\pi (m-1) \xi_{i} \sin\theta}, \\
		\left[ \steer_{y, i}(\theta, \varphi)\right]_{n}  &= e^{ - \bar{\jmath} 2\pi (n-1) \xi_{i} \cos\theta\sin\varphi},
	\end{align} 
	where $m\in\left\lbrace1, \dots, M_{i,z} \right\rbrace $ and $n\in\left\lbrace1, \dots, M_{i,y} \right\rbrace $. Note that the power of each path is normalized to simplify the analysis. 
	
	
	Based on the DL channel model expressed in (\ref{channel}), we discretize $ \theta $ and $ \varphi$ into $ \{ \theta(k) \}_{k=1}^{M_{i}} $ and $ \{ \varphi(k) \}_{k=1}^{M_{i}} $, respectively, where $ \theta(k) \in \left[ \theta_{\rm min}, \theta_{\rm max} \right] $ and $ \varphi(k) \in \left[ \varphi_{\rm min}, \varphi_{\rm max} \right] $. Let 
	\begin{align}
		\mathbf{A}_{i} &= \left[ \steer_{i}\left( \theta(1), \varphi(1) \right) , \dots, \steer_{i}\left( \theta(M_{i}), \varphi(M_{i}) \right)   \right] , \\
		\bm{\beta}_{\bm{q}, i} &= \left[  \beta_{\bm{q}, i}(\theta(1), \varphi(1)), \dots, \beta_{\bm{q}, i}(\theta(M_{i}), \varphi(M_{i})) \right]^{T},
	\end{align}
	then, the channel can be approximated by 
	\begin{equation}
		\mathbf{h}_{\bm{q}, i} = \mathbf{A}_{i} \bm{\beta}_{\bm{q}, i} .
	\end{equation}
	We assume that the complex-valued channel gains in different angles are uncorrelated, which means that 
	\begin{equation}
		\mean\left\lbrace   \beta_{\bm{q}, i}(\theta(k), \varphi(k)) \beta_{\bm{q}, i}^{*}(\theta(k'), \varphi(k'))  \right\rbrace  = S_{\bm{q}, i}(k) \delta( k-k'),
	\end{equation} 
	where sequence $ \{S_{\bm{q}, i}(k)\}_{k=1}^{M_{i}} \triangleq \mathcal{S}^{i}_{\bm{q}}$ describes the power distribution in the beam domain \cite{s}. Then, the channel covariance matrix related to the UT-$\bm{q}$ at frequency $f_{i}$ can be expressed as 
	\begin{align}
		\rr_{i}(\bm{q}) &= \mean\left\lbrace \mathbf{h}_{\bm{q}, i} \mathbf{h}_{\bm{q}, i}^{H} \right\rbrace \nonumber \\
		         &= \mathbf{A}_{i} \widetilde{\mathbf{R}}_{i}(\bm{q}) \mathbf{A}_{i}^{H} \nonumber \\
		         &= \mathbf{A}_{i} {\rm diag} \{ \mathcal{S}^{i}_{\bm{q}}\} \mathbf{A}_{i}^{H}.	\label{trans}	
	\end{align}
	 It is worth noting that (\ref{trans}) transforms the spatial covariance matrix into $\widetilde{\mathbf{R}}_{i}(\bm{q})={\rm diag} \{ \mathcal{S}^{i}_{\bm{q}}\} $, a diagonal matrix that reveals the channel power distribution in the beam domain \cite{reuse}. 
	
	Then, we consider the properties of matrix $\mathbf{A}_{i} $ for further analysis. It has been proved in \cite{dfts} that
	 when $M_{i}$ is sufficiently large, there exists a one-to-one approximation between $\theta(k)$ and $k$, i.e., $\sin\theta(k) \to \frac{2k}{M_{i}  }-1$ \cite{dfts}, so as the relationship between $\varphi(k)$ and $k$. Therefore, if $d_{i} = \lambda_{i} / 2$, elements in $\mathbf{A}_{i}$ can be further denoted as
	\begin{align}
	\left[ \mathbf{A}_{i} \right]_{m,n} 
		&= \displaystyle{
			e^{ - \bar{\jmath} 2 \pi \frac{ \left( m^{'}-1\right)  \left( n-1- \frac{M_{i}}{2}\right)}  {M_{i}} }     
		      e^{ - \bar{\jmath} 2 \pi \frac{ \left( m^{''}-1\right)  \left( n-1-\frac{M_{i}}{2}\right) } { M_{i}} }
		      } \nonumber \\
		&= \displaystyle{
			e^{- \bar{\jmath} 2 \pi \frac{ \left[  \left( m^{'}-1\right)  \left( n-1-\frac{M_{i}}{2}\right)  \right]  + \left[  \left( m^{''}-1\right)  \left( n-1-\frac{M_{i}}{2}\right)  \right]  } { M_{i}} } },
		\label{dft}
	\end{align}
	where $m^{'} = \lceil \frac{m}{M_{i,y}} \rceil$, $m^{''} = (m) {\rm mod} (M_{i,y}) $. 
	
	The result in (\ref{dft}) is critical because it indicates that the matrix $\mathbf{A}^{i}_{\bm{q}} $ tends to be the unitary two-dimensional discrete Fourier transform (2D-DFT) matrix when $M_{i}$ is sufficiently large. Similar approximations were derived for uniform linear arrays (ULA) in \cite{reuse, theta, dfts, ttjj}. Consequently, $\mathbf{A}_{i}$ is no longer an unknown matrix related to $f_{i}$, but rather a 2D-DFT matrix with its dimension solely depending on $M_{i}$. In this case, (\ref{trans})
	can be regarded as the eigenvalue decomposition (EVD) of $\rr_{i}(\bm{q})$ where columns in $\mathbf{A}_{i}$ are the eigenvectors and $\{S_{\bm{q}, i}(k)\}_{k=1}^{M_{i}}$ are the corresponding eigenvalues. That is to say, we can recover the spatial covariance matrix $\rr_{i}(\bm{q})$ as long as $\{S_{\bm{q}, i}(k)\}_{k=1}^{M_{i}}$ have been estimated. To conclude, the relationship between $\rr_{i}(\bm{q})$ related to UT-$\bm{q}$ at frequency $f_{i}$ and the PAS dependent upon eigenvalues is well established in multi-band massive MIMO systems.
	  
	\subsection{Channel Fingerprints Model}\label{cfmodel}	
		Based on the proposed channel model in \secref{channelmodel}, we now investigate the CF model for multi-band massive MIMO communication.

\begin{figure}[htbp]
		\centering
		\includegraphics[width=1\linewidth]{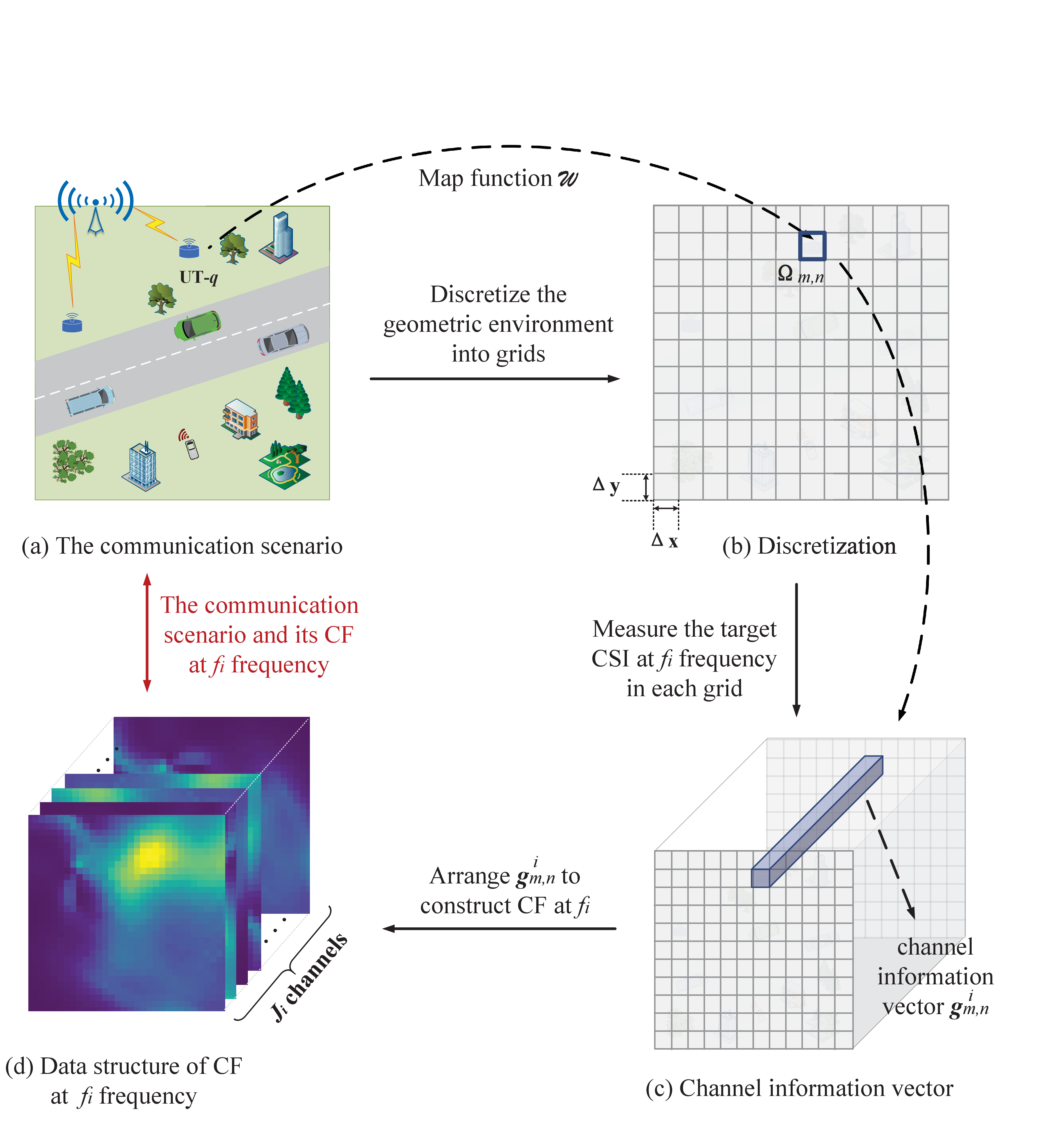}		
		\caption{CF model for multi-band massive MIMO transmission. (a) is the original communication scenario which is discretized into grids in (b). (c) provides the schematic diagrams of the measured channel information vector $\mathbf{g}^{i}_{m,n} $ and (d) presents the data structure of CF at frequency $f_{i}$.}
		\label{grid}
	\end{figure}

	As illustrated in \figref{grid}, the continuous geometric environment with the size $X \times Y$ is firstly discretized into grids, the set of which is denoted as $\Omega$. We define the length of each grid as $\Delta x$ and the width as $\Delta y$, then there are total $L_{x} \times L_{y}$ elements in $\Omega$, where $L_{x} = X / \Delta x$ and $L_{y} = Y / \Delta y $ are the numbers of columns and rows, respectively. For any UT located at $\bm{q} \in \mathcal{Q}$, it will be categorized into the corresponding $(m,n)$-th grid $\Omega_{m,n} \in \Omega$ so that we can obtain the channel information of UT-$\bm{q}$ with the CSI in grid $\Omega_{m,n}$.
	\vspace{-8mm}
	
	In order to measure the target CSI at frequency $f_{i}$ in $\Omega_{m,n}$, a trajectory $\mathcal{T}$ is randomly established in the grid with a receiver moving along it. Then, $T$ time slots $\{t_{1}, t_{2}, ..., t_{T}\}$ are selected to measure $\mathcal{H} = \{h_{m, n}(t_{i})\}_{i=1}^{T}$. Note that $\mathcal{T}$ is short enough that it can be approximately considered as a point in $\Omega_{m,n}$ relative to the entire region, so it is possible to describe the statistical CSI at $f_{i}$ in $\Omega_{m,n}$ through the calculation of $\mathcal{H}$. Assume that the statistical CSI comprises a total of $J_{i}$ elements, then the grid $\Omega_{m,n}$ would correspond to a channel information vector $\mathbf{g}^{i}_{m,n} \in \mathbb{C}^{J_{i}\times 1}$.
	By arranging $\mathbf{g}^{i}_{m,n}$ in the order of grid rows and columns, we can constitute the channel fingerprints at frequency $f_{i}$, which is denoted as $\bm{\mathcal{G}}_{i} \in \mathbb{C}^{L_{y}\times L_{x}\times J_{i}}$. Suppose $\bm{\mathcal{G}}_{i}$ are constructed in all frequencies, we would then obtain CF for this multi-band massive MIMO system, which is denoted as $\bm{\mathcal{G}} = \{\bm{\mathcal{G}}_{1}, \bm{\mathcal{G}}_{2}, \dots, \bm{\mathcal{G}}_{N}\}$.

    It is evident that the three-dimensional matrix $\bm{\mathcal{G}}_{i}$ is structurally similar to an image with the number of its channels being $J_{i}$. However, it is crucial to emphasize that the elements in $\bm{\mathcal{G}}_{i}$ correspond to the grids instead of the pixels, even though $\bm{\mathcal{G}}_{i}$ shares a similar structure with multichannel images. To model $\bm{\mathcal{G}}_{i}$ as an image, a grid-to-pixel conversion must be performed, where each grid contains (or corresponds to) multiple pixels. More details will be explained in \secref{pf}.
    
	\subsection{Problem Formulation}\label{pf}
	Based on the channel model and CF model discussed above, we next formulate the CF extrapolation problem for multi-band massive MIMO transmission.
	
	First, we have to determine the target channel information in CF.
	With respect to the $(m,n)$-th grid $\Omega_{m,n}$ in the target region, Suppose that we vectorize $\rr_{i}(m,n)= \mathbf{A}_{i} {\rm diag} \{ \mathcal{S}^{i}_{m,n}\} \mathbf{A}_{i}^{H} \in\mathbb{C}^{M_{i}\times M_{i}}$ to construct the channel information vector $\mathbf{g}^{i}_{m,n}$, which means that there are totally $J_{i} = M_{i} \times M_{i}$ elements in $\mathbf{g}^{i}_{m,n} $. In that case, the storage overhead and computational cost required for the construction are immense since $J_{i}$ is excessive with the antenna array expanding dramatically in multi-band massive MIMO systems. In fact, discussions in \secref{channelmodel} have pointed out that $A_{i}$ here is a known 2D-DFT matrix so that storing the eigenvalues $\{S_{m,n,i}(k)\}_{k=1}^{M_{i}}$ is sufficient to obtain $\rr_{i}(m,n)$, reducing $J_{i}$ to $M_{i}$ and decreasing the storage cost while maximally preserving the characteristics of statistical CSI. Consequently, the channel information vector $\mathbf{g}^{i}_{m,n} $ can be written as
	\begin{equation}
		\mathbf{g}^{i}_{m,n} = [ S_{m,n,i}(1), S_{m,n,i}(2), \dots, S_{m,n,i}(M_{i})]^{T} \in \mathbb{R}^{J_{i}\times 1},
	\end{equation}
	where $J_{i}=M_{i}$. According to the CF model in \secref{cfmodel}, we arrange $\mathbf{g}^{i}_{m,n}$ sequentially and denote this constructed CF at $f_{i}$ with eigenvalues $\{S_{m,n,i}(k)\}_{k=1}^{M_{i}}$ as $\bm{\mathcal{G}}_{i}(S)$. 
	
	Next, we need to perform the grid-to-pixel conversion to formulate $\bm{\mathcal{G}}_{i}(S)$ as an multichannel image. A straightforward yet efficient way is to make each grid uniformly cover $\rho \times \varrho$ sub-grids, with each sub-grid having the same channel information as the corresponding grid. Then, we model these sub-grids as pixels, where $\Delta x / \rho$ and $\Delta y / \varrho$ are resolutions of the sub-grids (or pixels) along the $x$-axis and $y$-axis, respectively. As a result, the matrix $\bm{\mathcal{G}}_{i}(S) \in \mathbb{C}^{L_{y}\times L_{x}\times M_{i}}$ is converted to a multichannel image $\bm{\mathcal{I}}_{i}(S) \in \mathbb{C}^{\varrho L_{y}\times \rho L_{x}\times M_{i}}$, with the $(h, w, m_{i})$-th element of $\bm{\mathcal{I}}_{i}(S)$ expressed as
		\begin{equation}
			[\bm{\mathcal{I}}_{i}(S)]_{h,w,m_{i}} = [\bm{\mathcal{G}}_{i}(S)]_{ \lfloor\frac{h}{\varrho}\rfloor, \lfloor \frac{w}{\rho} \rfloor, m_{i} }, \label{pii}
	\end{equation}
    where $h=1,2,\dots,\varrho L_{y}$, $w=1,2,\dots,\rho L_{x}$, and $m_{i}=1,2,\dots,M_{i}$. When $\rho=\varrho=1$, the concepts of grids and pixels become the same, aligning with the modeling approach adopted by the majority of current researches. In fact, the grid-to-pixel conversion should not be neglected because it can enable the refining scheme, which will be explained in \secref{rs}. After the operation in (\ref{pii}), we translate CF at $f_{i}$ into the image $\bm{\mathcal{I}}_{i}(S)$, with the number of its channels being $M_{i}$. 
	
	Since all frequencies in multi-band massive MIMO systems work in the same wireless scenario where the geometry of the environment and its propagation characteristics like blockage are consistent \cite{sub2mmw}, there exists a specific correlation among CF at different frequencies. So it is possible for us to explore the mapping function $\Psi_{i,j}$ between $\bm{\mathcal{I}}_{i}(S)$ and $\bm{\mathcal{I}}_{j}(S)$, which is denoted as 
	\begin{equation}
		\Psi_{i,j} : \bm{\mathcal{I}}_{i}(S) \to \bm{\mathcal{I}}_{j}(S).
		\label{sx}
	\end{equation}
	Note that (\ref{sx}) implies a bidirectional relationship between $i$ and $j$ since $i, j$ can be any as long as they are not the same. This is crucial for extrapolation tasks, as there exists a bijective relationship between CF at different frequencies.
    
     In practice, it is challenging to derive an analytical solution using traditional extrapolation methods, which motivates us to employ machine learning (ML) to explore the bidirectional mapping relationship. Since $\bm{\mathcal{G}}_{i}(S)$ has been transformed into the multichannel image $\bm{\mathcal{I}}_{i}(S)$, the problem of CF extrapolation for multi-band massive MIMO transmission is modeled as an image translation task, whose objective is to realize the conversion between different data domains. Hence, the problem can be expressed as
    \begin{align}
    	\mathop{\arg \min}\limits_{\{\Theta, \Theta'\}} \Big\{ &\mean \left[   \| \Psi_{i,j}[\bm{\mathcal{I}}_{i}(S); \Theta] -\bm{\mathcal{I}}_{j}(S) \|^{2}_{F} \right]  \nonumber \\
    	& + \mean \left[ \| \pn_{j,i}[\bm{\mathcal{I}}_{j}(S); \Theta'] - \bm{\mathcal{I}}_{i}(S) \|^{2}_{F} \right]  \Big\} ,
    \end{align}
    where $\Theta$ and $\Theta'$ are sets of the trainable parameters, and $\pn_{j,i}$ is used to approximate the inverse mapping of $\Psi_{i,j}$, thus ensuring that the reciprocal relationship of (\ref{sx}) can be optimally fitted.

	\section{Channel Fingerprints Extrapolation Based On the Cycle-Consistent Generative Networks } \label{CU}
 Since the CF extrapolation problem has been formulated as an image translation task in CV, we adopt the generative model, a practical approach for image-to-image (I2I) problems, to achieve feature extraction and mapping fitting. Therefore, in this section, we propose a cycle-consistent generative networks-based extrapolation approach to predict CF for multi-band massive MIMO transmission, called CF-CGN. We begin with the illustration of the network structure. Then, a cycle consistency loss is designed to couple the networks together, equipped with a weight function that aligns with iterative optimization steps. Afterwards, we prepare the data for network training and propose a joint parameter update method. Finally, a refining scheme is introduced to improve the extrapolation accuracy.

	\subsection{Network Architecture}
	\begin{figure*}[t!]
		\centering
		\includegraphics[width=1\textwidth]{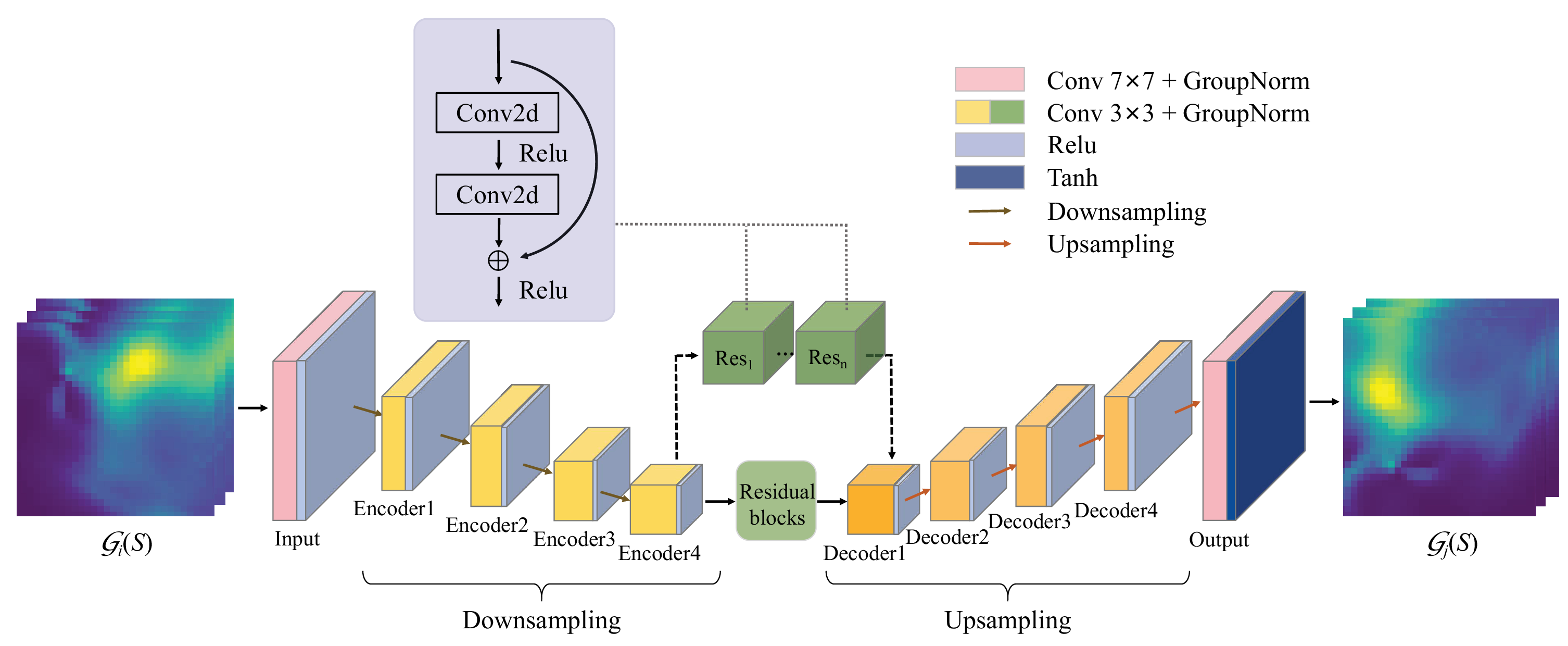}		
		\caption{Architecture of the proposed generative network for channel fingerprints extrapolation in multi-band massive MIMO systems.}
		\label{net}
	\end{figure*}
	
	CF-CGN is similar to an ``image transformer'': it compresses the raw data into a latent space by extracting pivotal features from the input and then decodes the data manifold to the output based on unique characteristics of the target CF.
    The network architecture is shown in \figref{net}, which can be divided into five parts: input module, downsampling module, residual blocks, upsampling module, and output module.	
	\subsubsection{Input Module}
	One convolutional layer with a rectified linear unit (ReLU) is adopted as the input module, which is designed to reshape the data and initially extract features with a large receptive field. Denote the input and output as $\bm{\mathcal{I}}_{i}(S) \in \mathbb{C}^{\varrho L_{y}\times \rho L_{x}\times M_{i}}$ and $\bm{\mathcal{X}} \in \mathbb{R}^{H\times W\times P}$, respectively, we have
	\begin{equation}
		\bm{\mathcal{X}} = {\rm Conv2}(\bm{\mathcal{I}}_{i}(S)),
	\end{equation}
	where ${\rm Conv2}(\cdot)$ represents the operation of image convolution.
	
		
	\subsubsection{Downsampling Module}
	The downsampling module consists of four convolutional layers to extract features of $\bm{\mathcal{X}}$ further, with the kernel size smaller than that in the input module. To facilitate the following analysis, we denote the outcome as $\bm{\mathcal{X}'} = {\rm Conv2}(\bm{\mathcal{X}} )$. Next, a group normalization (GN) is introduced to standardize data in groups. Define the output of GN as $\bm{\mathcal{X}}''$, the $g$-th group of $\bm{\mathcal{X}}''$ can be written as
	\begin{equation}
		[\bm{\mathcal{X}}'']_{g} = [{\rm GN}(\bm{\mathcal{X}'})]_{g} = \gamma_{g}     \dfrac{\bm{\mathcal{X}}_{g}' - \mu_{g}}{\sqrt{\sigma_{g}^{2} + \epsilon}}    + b_{g},
		\label{gn}
	\end{equation}
	where $\bm{\mathcal{X}}_{g}'$ is the $g$-th group of $\bm{\mathcal{X}}'$ and $\mu_{g}, \sigma_{g}$ represent its mean and variance, respectively. $\gamma_{g}, b_{g}$ denote the learnable parameters and $\epsilon$ is a tiny constant used to enhance stability. Note that GN is not dependent on the batch size so that the performance will be more stable during training. Afterwards, the ReLU is followed to enhance generalization by introducing nonlinear operation. 
	To conclude, the final output of the downsampling module is given by
	\begin{equation}
		\bm{\mathcal{X}}_{1} = {\rm ReLU}({\rm GN}({\rm Conv2(\bm{\mathcal{X}})})).
	\end{equation}

	The essence of the downsampling module is to extract features from the input and then compress them to the latent space $\mathcal{Z}$. This process is similar to the encoding operation since it reduces the data dimension and gets the latent representation of $\bm{\mathcal{X}}$, laying the foundation for CF extrapolation.
		
	\subsubsection{Residual Blocks}
	To prevent the issues of gradient vanishing and gradient explosion, we introduce residual blocks to enhance the network performance \cite{res}. As shown in \figref{net}, each residual block comprises two $3 \times 3$ convolutional layers with the stride set to be $1$. By stacking several blocks, we can deepen the network to extract more detailed features in $\bm{\mathcal{I}}_{i}(S) $ while avoiding the network degradation.
		
	\subsubsection{Upsampling Module}
	In correspondence with the downsampling module, the upsampling module is also composed of four convolutional layers with each layer equipped with the GN and ReLU. Therefore, the output is given by
	\begin{equation}
			\bm{\mathcal{X}}_{3} = {\rm ReLU}({\rm GN}({\rm TransConv2}(\bm{\mathcal{X}}_{2} ))),
	\end{equation}  
	where $\bm{\mathcal{X}}_{2}$ is the output of residual blocks and ${\rm TransConv2}(\cdot)$ is the transposed convolution with out-padding set to be $1$ and stride to be $2$.	
	The core of the upsampling module lies in reconstructing the target domain $\{\bm{\mathcal{X}}_{3}\}$ from latent representation of $\{\bm{\mathcal{X}}_{2}\} \subseteq  \mathcal{Z}$ based on the correlation between their features, which is very similar to the decoding operation.

	\subsubsection{Output Module}
	The output module aims to recover $\bm{\mathcal{I}}_{j}(S)$ from $\{\bm{\mathcal{X}}_{3}\}$ by a few convolutional layers. Suppose that there are totally $M_{j}$ kernels with the size being $(K_{1} \times K_{2} \times C' )$, then, we can express the output as $\bm{\mathcal{X}}_{4} = {\rm Conv2}(\bm{\mathcal{X}}_{3}) \in \mathbb{C}^{\varrho L_{y}\times \rho L_{x}\times M_{j}}$. It is worth noting that we adopt the hyperbolic tangent function (Tanh) here instead of ReLU for activation to keep the data within $[-1, 1]$, which is consistent with the format of processed input data.
	
		
	To conclude, the whole network can learn the latent representation of $\bm{\mathcal{I}}_{i}(S)$ and then extrapolate $\bm{\mathcal{I}}_{j}(S)$ accordingly. As a result, we adopt the basic loss as
		\begin{equation}
			\mathcal{L}_{basic}(\Psi_{i,j}) = \mean \{ \| \Psi_{i,j}(\bm{\mathcal{I}}_{i}(S)) - \bm{\mathcal{I}}_{j}(S) \|_{F}^{2} \}. \label{basic}
		\end{equation}

	\subsection{Cycle Consistency Loss with Variable Weight} \label{fw}
    Cycle consistency loss is a fundamental component in image translation and domain adaptation models, ensuring that images transformed from one domain to another can be translated back to their original domain with minimal distortion \cite{cyclegan}. Its robustness, combined with its ability to enforce the bidirectional consistency, has enabled widespread applications in I2I translation \cite{i2}, voice conversion \cite{vo}, and synthetic aperture radar (SAR) image reconstruction \cite{cy}. As a result, we will next design the cycle consistency loss to achieve the CF's bidirectional extrapolation as expressed in (\ref{sx}).
    
	We first adopt a pair of generative networks described in \figref{net}, which are denoted as 
		\begin{equation}
			\begin{cases}
				\Psi_{i,j}:&\bm{\mathcal{I}}_{i}(S) \to \bm{\mathcal{I}}_{j}(S), \\
				\Psi_{j,i}:&\bm{\mathcal{I}}_{j}(S) \to \bm{\mathcal{I}}_{i}(S). \label{p2}
			\end{cases}
		\end{equation}
		
	\textit{Remark 1:} Define a pair of domains $\bm{\mathcal{D}}_{i}$ and $\bm{\mathcal{D}}_{j}$, if two generative networks are just trained independently to fit the following mapping,
		\begin{equation}
			\begin{cases}
				\Gamma_{i\to j}(\bm{d}_{i})= \bm{d}_{j}, \\
				\Gamma_{j\to i}(\bm{d}_{j}^{'})= \bm{d}_{i}',
			\end{cases}
		\end{equation} 
	where $\bm{d}_{i}, \bm{d}_{i}' \in \bm{\mathcal{D}}_{i}$ and $\bm{d}_{j}, \bm{d}_{j}' \in \bm{\mathcal{D}}_{j}$, then the equality of $\bm{d}_{j}$ and $\bm{d}_{j}'$ does not guarantee the equality of $\bm{d}_{i}$ and $\bm{d}_{i}^{'}$ since these mappings are highly under-constrained. 
		
\begin{figure}[htbp]
		\centering
		\includegraphics[width=1\linewidth]{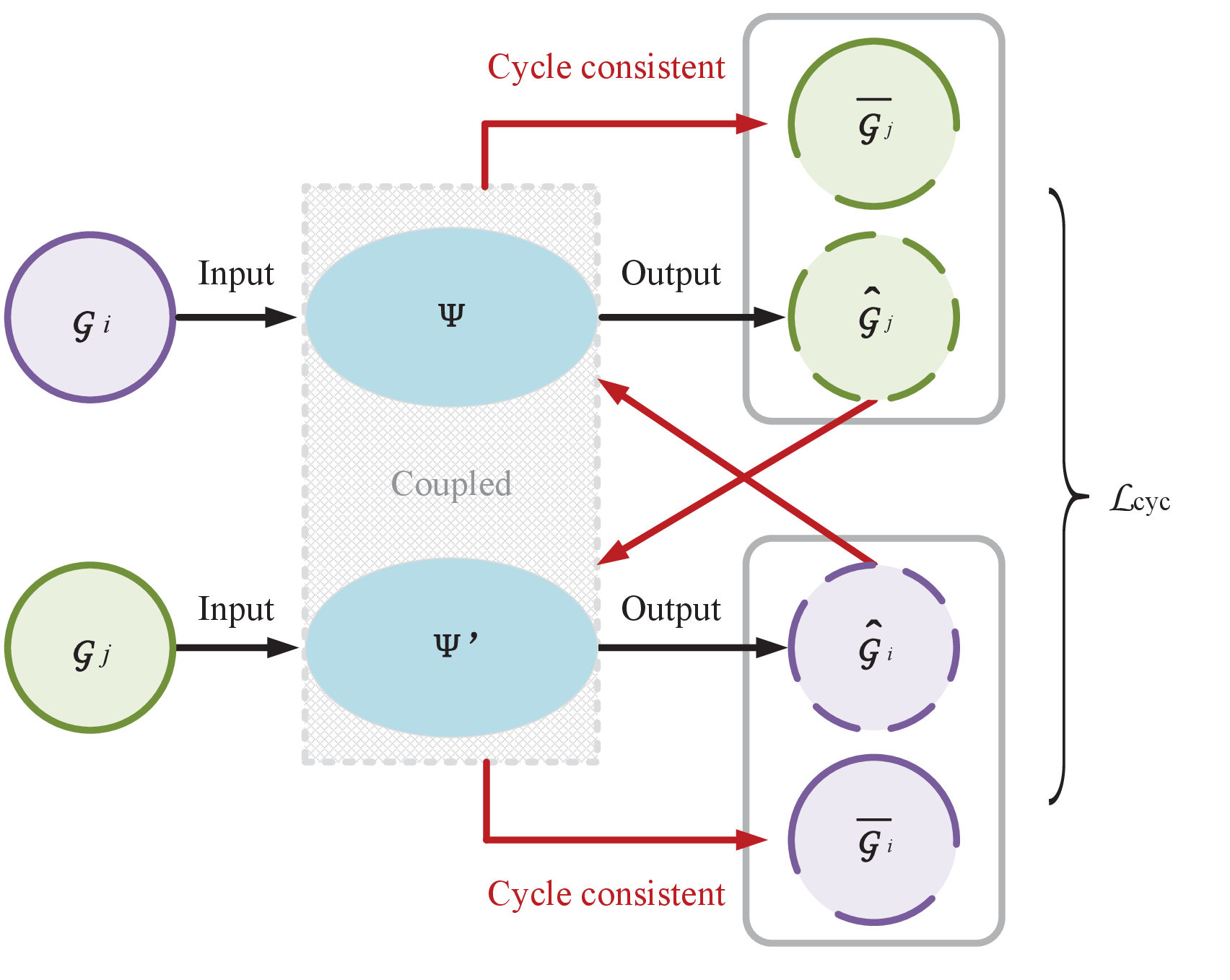}		
		\caption{Schematic diagram of the cycle consistency loss.}
		\label{cyc}
	\end{figure}

	Concerning CF extrapolation for multi-band massive MIMO transmission, this remark reveals that although both generative networks in (\ref{p2}) can respectively achieve the mapping from $\bm{\mathcal{I}}_{i}(S)$ to $\bm{\mathcal{I}}_{j}(S)$ and $\bm{\mathcal{I}}_{j}(S)$ to $\bm{\mathcal{I}}_{i}(S)$, they can not form a bidirectional mapping in (\ref{sx}), that is, they fail to constitute reciprocal functions. Consequently, we introduce an additional requirement \cite{cyclegan} during the training process to emulate the reciprocal relationship, which is given by 
	\begin{equation}
		\begin{cases}
			\Psi_{i,j} \left[  \pn_{j,i} [\bm{\mathcal{I}}_{j}(S); \Theta']; \Theta \right]  = \bm{\mathcal{I}}_{j}(S), \\
			\pn_{j,i} \left[ \Psi_{i,j} [ \bm{\mathcal{I}}_{i}(S); \Theta]; \Theta' \right]  = \bm{\mathcal{I}}_{i}(S).
		\end{cases}
		\label{re}
	\end{equation}
	According to (\ref{re}), we tailor cycle consistency losses to make sure that the paired networks are ``cycle-consistent'', as shown in \figref{cyc}, which can be expressed as
	\begin{align}
		\mathcal{L}_{cyc} (\Psi_{i,j}, \pn_{j,i})= &\mean \{ \| \Psi_{i,j}( \pn_{j,i}(\bm{\mathcal{I}}_{j}(S)) ) - \bm{\mathcal{I}}_{j}(S) \|^{2}_{F} \} \nonumber \\
		&+ \mean \{ \| \pn_{j,i}( \Psi_{i,j}(\bm{\mathcal{I}}_{i}(S)) ) - \bm{\mathcal{I}}_{i}(S) \|^{2}_{F} \}. \label{cycloss}
	\end{align}
		
	\textit{Remark 2:} If the cycle consistency loss is applied, the paired networks $\Psi_{i,j}$ and $\pn_{j,i}$ will then become reciprocal functions and satisfy the following properties,
	\begin{equation}
		\begin{cases}
			\Psi_{i,j} \circ \pn_{j,i} = I_{i\to i}, \\
			\pn_{j,i} \circ \Psi_{i,j} = I_{j\to j},
		\end{cases}
	\end{equation}
    where $I_{i\to i}$ and $I_{j\to j}$ are two identity mappings. 
		 
    This remark emphasizes the relationship between $\Psi_{i,j}$ and $\pn_{j,i}$, that is, the composition of our proposed paired networks is similar to a set of auto-encoders, a classical model in generative AI. However, these auto-encoders have specific internal structures: they reconstruct the data through an intermediate representation, which is another domain translated from this data instead of the latent space.      
	
	Since the cycle consistency loss is of great importance in the training process, we need to design an effective weight to control its influence and balance two loss functions $\mathcal{L}_{basic}$ and $\mathcal{L}_{cyc}$.	
	It is well known that the proposed network will continuously optimize its parameters in the training process by gradient descent. The cycle consistency loss, which aims to construct the reciprocal functions, is thus expected to adjust its weight as the network updates automatically. One possible way is to design the weight as a function $f(\omega) \in [0, 1]$ with the normalized iterative index $\omega = d/D \in [0, 1]$ being the independent variable, where $d=1, 2, \dots, D$ and $D$ is the total number of iterations. 
	
	To conclude, we design $\mathcal{L}_{cyc}$ to fit the reciprocal relationship and propose a variable weight function $f(\omega)$ to make $\mathcal{L}_{cyc}$ change adaptively with the iterative process. As a result, the weighted cycle consistency loss is given by
	\begin{equation}
		\mathcal{L}_{wcyc}(\Psi_{i,j}, \pn_{j,i}) = f(\omega)\mathcal{L}_{cyc}(\Psi_{i,j}, \pn_{j,i}). \label{full}
	\end{equation}  

	\subsection{Network Training}
	Next, we introduce the training process of CF-CGN, including the data preparation and the joint training strategy. 
		
	\subsubsection{Data Preparation}
	We adopt the ``max-min'' linear normalization to scale the raw CF data into $[0, 1]$, which is beneficial to the stability of network training. Denote $[\bm{\mathcal{I}}_{i}(S)]_{m_{i}}$ as the $m_{i}$-th channel of $\bm{\mathcal{I}}_{i}(S)$, the $(h, w, m_{i})$-th element of the normalized data is given by
	\begin{equation}
		[\bm{\mathcal{I}}_{i}^{'}(S)]_{h, w, m_{i}} = \dfrac{ [\bm{\mathcal{I}}_{i}(S)]_{h,w, m_{i}} - \{ [\bm{\mathcal{I}}_{i}(S)]_{m_{i}} \}_{\rm min}  }{ \{ [\bm{\mathcal{I}}_{i}(S)]_{m_{i}} \}_{\rm max} - \{ [\bm{\mathcal{I}}_{i}(S)]_{m_{i}} \}_{\rm min}    }. \label{mm}
	\end{equation} 

	\subsubsection{Joint Training Strategy}
	As mentioned in \textit{Remark 1}, training paired networks independently cannot forge them to be reciprocal functions. Consequently, we design a joint training strategy with cycle consistency losses to optimize two sets of parameters synchronously, thus achieving the bidirectional mapping. Specifically, the networks are trained to minimize the joint loss $\mathcal{L}_{joint}$ and updated through Adam with the learning rate being $\alpha$. Based on (\ref{basic}) and (\ref{full}), the joint loss can be expressed as
	\begin{equation}
		\mathcal{L}_{joint} = \mathcal{L}_{basic}(\Psi_{i,j}) + \mathcal{L}_{basic}(\pn_{j,i}) + \mathcal{L}_{wcyc}(\Psi_{i,j}, \pn_{j,i}),
		\label{jo}
	\end{equation}
	where $\mathcal{L}_{wcyc}(\Psi_{i,j}, \pn_{j,i})$ serves as a bridge to couple two networks together. 
	In the actual training process, we test the designed weight function $f(\omega)$ from the three categories for simplicity: monotonically increasing functions $\mathcal{F}_{ic}$, monotonically decreasing functions $\mathcal{F}_{dc}$, and constant functions $\mathcal{F}_{cn}$. With respect to each category, several typical functions will be selected for comparison. 
	
	To conclude, we can extrapolate CF based on the prior information at any frequency for multi-band massive MIMO systems with the training process mentioned above, which provides novel insights into the utilization and integration of unlicensed spectra.  
		
	\subsection{Refining Scheme} \label{rs}
	Consider the trained network $\Psi_{i,j}$, we can get the estimated CF by $\bar{\bm{\mathcal{I}}}_{j}(S) = \Psi_{i,j}(\bm{\mathcal{I}}_{i}(S); \Theta)$, where $\bar{\bm{\mathcal{I}}}_{j}(S) \in \mathbb{C}^{\varrho L_{y}  \times \rho L_{x} \times M_{j}}$. Due to the grid-to-pixel conversion discussed in \secref{pf}, $\bm{\mathcal{I}}_{j}(S)$ represents the image format of CF, so we have to perform pixel-to-grid operation to restore $\gjs$.
	Note that each gird in $\gjs$ has been divided into $\rho \times \varrho$ sub-grids (pixels) in (\ref{pii}), with each sub-grid (pixel) possessing the same channel information as the corresponding grid. Hence, we adopt an average pooling to recover $\gjs$, where the value of each grid in $\gjs$ corresponds to an average of $\rho \times \varrho$ sub-grids (pixels) in $\bar{\bm{\mathcal{I}}}_{j}(S)$. Define the size of the average pooling kernel as $\rho \times \varrho$, the $(l_{y}, l_{x}, m_{j})$-th element of $\gjs$ can be expressed as
	\begin{equation}
		[\bm{\mathcal{G}}_{j}(S)]_{l_{y}, l_{x}, m_{j}} = \dfrac{1}{\rho\varrho} \sum_{\imath=0}^{\varrho-1} \sum_{\ell=0}^{\rho-1} [\bm{\mathcal{I}}_{j}(S)]_{l_{y}\varrho-\imath, l_{x}\rho-\ell, m_{j}},
	\end{equation}
	with the size of $\bm{\mathcal{G}}_{j}(S)$ being $(L_{y} , L_{x} ,M_{j})$.
	
    In contrast to the methodology adopted in the majority of researches where one grid equals to one pixel ($\rho=\varrho=1$), first conducting grid-to-pixel conversion described in \secref{pf} and then using average pooling for restoration can refine the outcomes and enhance the accuracy of CF extrapolation. In fact, the essence of grid-to-pixel conversion lies in replicating a single grid into $\rho\varrho$ copies so that the network can conduct multiple parallel predictions simultaneously and then integrate the results. Although extrapolation errors in neural networks are inevitable, this strategy can effectively mitigate the impact of extreme errors, reduce the variance of the model, and enhance its generalization capability. This concept bears a resemblance to the neural network ensemble methods.

	\section{Numerical Results} \label{nr}
	In this section, we provide numerical results to evaluate the performance of our proposed CF-CGN. We begin with the dataset generation and the experiment setup. Then, we explore the influence of variable weight function $f(\omega)$. Compared with the linear extrapolation-based method and CycleGAN-based approach, we conduct a series of experiments under different scenarios to demonstrate our method's accuracy and generalization ability. Finally, the sum rate performance for CF-assisted multi-band massive MIMO transmission is presented, together with the comparison of time and storage complexities.
	
	\subsection{Datasets}
	The datasets are generated by QuaDRiGa \cite{quad}, which follows the geometry-based stochastic channel modeling approach to simulate realistic radio channel impulse responses for mobile radio networks. We adopt the communication scenarios with propagation conditions including both line of sight (LOS) and non-line of sight (NLOS). For each scenario, a target region $\mathcal{A}$ with the size of $32 \times 32 \ {\rm m}^{2}$ is designated, and we discretize it into grids where $\Delta x, \Delta y$ are both set to be $2 \ {\rm m}$. The trajectory $\mathcal{T}$ is set to be circular for measurement in each grid, with $364$ time slots taken along the path based on \secref{cfmodel}.
	
	To facilitate the conduct of our experiments, we adopt two frequencies $\fmin$ and $\fmax$ as an example to verify the proposed method. In case of no special mention, $f_{1}$ is set to be $2$ GHz and $f_{2}$ to be $5$ GHz, a typical frequency of Wi-Fi in the unlicensed spectra. Firstly, the UPA with $M_{1} = 8 \times 8$ antennas, as well as the other co-located UPA with $M_{2} = 10 \times 10$ antennas, is arranged at the BS, which is randomly located in $\mathcal{A}$. Then, for each grid in $\mathcal{A}$, we get channels of $\fmin$ and $\fmax$ in $364$ time slots and calculate the channel covariance matrix according to (\ref{trans}). Next, we conduct the eigenvalue decomposition and rearrange the eigenvalues in descending order to construct $\bm{g}_{m,n}^{1}$ and $\bm{g}_{m,n}^{2}$. After that, we arrange $\bm{g}_{m,n}^{1}$ sequentially to construct $\bm{\mathcal{G}}_{1}(S)$, with the same operation executed on $\bm{g}_{m,n}^{2}$ to construct $\bm{\mathcal{G}}_{2}(S)$. Finally, we set $\varrho = \rho = 2$ to achieve the grid-to-pixel conversion in (\ref{pii}), with resolutions of pixels along the $x-$axis and $y-$axis being $1$.
	All parameters used to generate datasets for the multi-band massive MIMO system are displayed in \tabref{tb:simulation parameters}.
	
	\subsection{Experiment Setup}
	 In correspondence with \figref{net}, the structure of CF-CGN is enumerated thoroughly in \tabref{tb2}, with the size of the input tensors being ($32 \times 32 \times 100$) and ($32 \times 32 \times 64$).  During the training process, we utilize the Adam optimizer and train 80 epochs where the learning rate is set to 0.0001 for the first 40 epochs and then linearly decays to zero over the following 40 epochs. The batch size is set to be 16. We conduct our simulations based on Pythorch and MATLAB, with the computer equipped with an Intel(R) Core(TM) i7-12700 and a Geforce GTX 4090. 
	 All hyper-parameters are listed in \tabref{tb:simulation parameters}.
	 
	 In order to evaluate the performance of CF-CGN, we employ the normalized mean square error (NMSE) as the evaluation criterion, which is defined as
	 \begin{equation}
	 	{\rm NMSE}{\rm (dB)} =\dfrac{1}{M} \sum_{m=1}^{M} \left\lbrace 10 \log \dfrac{\| [\hat{\bm{\mathcal{G}}}_{i}(S)]_{m} - [\gis]_{m} \|_{F}^{2}}{ \| [\gis]_{m} \|_{F}^{2} } \right\rbrace ,
	 \end{equation}
	 where $M\in\{M_{i}, M_{j}\}$ and $[\hat{\bm{\mathcal{G}}}_{i}(S)]_{m}$ is the $m$-th channel of the estimated CF.
	 
	 	\newcolumntype{L}{>{\hspace*{-\tabcolsep}}l}
	 \newcolumntype{R}{c<{\hspace*{-\tabcolsep}}}
	 \definecolor{lightblue}{rgb}{0.93,0.95,1.0}
	 \begin{table}[htbp]
	 	\captionsetup{font=footnotesize}
	 	\captionsetup{justification=centering}
	 	\caption{Parameters used to generate datasets and hyper-parameters used to train the networks}\label{tb:simulation parameters}
	 	\centering
	 	\ra{1.45}
	 	\scriptsize
	 	\begin{tabular}{LR}
	 		\toprule
	 		Parameter &  Value\\
	 		\midrule
	 		\rowcolor{lightblue}
	 		Communication scenario \ \ \ \ \ \ \ \ \ \ \ \ \ \ \ \ \ \ \ \ \ \ \ \ \ \  & LOS, NLOS   \\
	 		Size of the target region $\mathcal{A}$ & $32$ m $\times$ $32$ m \\
	 		\rowcolor{lightblue}
	 		Length of the grid $\Delta x$ & $2$ m \\
	 		Width of the grid $\Delta y$ & $2$ m \\
	 		\rowcolor{lightblue}
	 		Trajectory $\mathcal{T}$ & circle \\
	 		Number of time slots & $364$ \\
	 		\rowcolor{lightblue}
	 		Height of the BS & $10$ m   \\
	 		Frequency 1 & $2$ GHz\\
	 		\rowcolor{lightblue}
	 		Antenna array for $f_{1}$ & $8 \times 8$\\
	 		Frequency 2 & $3.9, 5, 24$ GHz\\
	 		\rowcolor{lightblue}
	 		Antenna array for $f_{2}$ &  $10 \times 10$ ,  $12 \times 12$ , $15 \times 15$ \\
	 		Height of UTs & $1.5$ m\\	
	 		\rowcolor{lightblue}		
	 		Power constraint & 1\\
	 		Epochs               & 80 \\
	 		\rowcolor{lightblue}
	 		Delay epochs  & 40 \\
	 		Learning rate   &   0.0001 \\
	 		\rowcolor{lightblue}
	 		Batch size    & 16 \\
	 		Optimizer  & Adam\\
	 		\bottomrule
	 	\end{tabular}
	 \end{table}
	 
	 \newcolumntype{L}{>{\hspace*{-\tabcolsep}}l}
	 \newcolumntype{R}{c<{\hspace*{-\tabcolsep}}}
	 \definecolor{lightblue}{rgb}{0.93,0.95,1.0}
	 \begin{table}[htbp]
	 	\captionsetup{font=footnotesize}
	 	\captionsetup{justification=centering}
	 	\caption{Parameters for CycGCFp structure}\label{tb2}
	 	\centering
	 	\ra{1.45}
	 	\scriptsize
	 	\begin{tabular}{LccR}
	 		\toprule	 		
	 		Layers \ \ \ \ \ \ \ \ & \ \ \ \ \ \  Kernel Size \ \ \ \ \ \ &\ \ \ \ \ \ Stride\ \ \ \ \ \ & Padding\\
	 		\midrule
	 		\rowcolor{lightblue}
	 		Input             & $7 \times 7$            & $1 \times 1$        & 3                    \\
	 		Encoder 1-4  & $3 \times 3$            & $2 \times 2$           & 1 \\
	 		\rowcolor{lightblue}
	 		Residual        & $3 \times 3$            & $1 \times 1$          & 1 \\
	 		Decoder 1-4  & $3 \times 3$            & $2 \times 2$         & 1 \\
	 		\rowcolor{lightblue}
	 		Output         & $7 \times 7$            & $1 \times 1$            & 3 \\
	 		\bottomrule
	 	\end{tabular}
	 \end{table}

	\newcolumntype{L}{>{\hspace*{-\tabcolsep}}l}
	\newcolumntype{R}{c<{\hspace*{-\tabcolsep}}}
	\definecolor{lightblue}{rgb}{0.93,0.95,1.0}
	\begin{table*}[t]
		\captionsetup{font=footnotesize}
		\captionsetup{justification=centering}
		\caption{NMSE performance of the CycGCFp-based channel fingerprints extrapolation under different types of $f(\omega)$} \label{fwb}
		\centering
		\ra{1.4}
		\scriptsize
		\begin{tabular}{LcccccccR}
			\toprule
			\multirow{2}{*}{}& \multicolumn{1}{c}{\multirow{2}{*}{$\mathcal{F}_{0}$ (without $\mathcal{L}_{cyc}$)}} & \multicolumn{3}{c}{$\mathcal{F}_{ic}$} & $\mathcal{F}_{cn}$  & \multicolumn{3}{c}{$\mathcal{F}_{dc}$} \\
			\cline{3-5} \cline{7-9}
			& & $f_{1} (\omega) = \omega$ & $f_{2} (\omega) = \omega^{3} $ & $f_{3} (\omega) = 1-e^{-3 \omega} $ & ($f_{0} (\omega) = 0.5$) & $f_{4} (\omega) = 1-\omega $ & $f_{5} (\omega) = 1-\omega^{3} $ & $ f_{6} (\omega) = e^{-3 \omega}$\\
			\midrule
			\rowcolor{lightblue}
			$f_{1}\to f_{2}$ &0.0352 &\textbf{0.0245} &0.0252 &0.0253 &0.0275 & 0.0261& 0.0256&0.0253\\
			$f_{2}\to f_{1}$ &0.0367 &\textbf{0.0256} &0.0265 &0.0263 &0.0269 & 0.0269& 0.0261&0.0266\\
			\bottomrule
		\end{tabular}
	\end{table*}

    \subsection{Influence of the Variable Weight Function $f(\omega)$}
	As described in \secref{fw}, variable weight function $f(\omega)$ can control the influence of cycle consistency loss $\mathcal{L}_{cyc}$ in the full object according to (\ref{jo}) and (\ref{full}). Therefore, we select three typical functions from $\mathcal{F}_{ic}$ and $\mathcal{F}_{dc}$ respectively to evaluate their effects. Specifically, $\mathcal{F}_{ic}$, as well as $\mathcal{F}_{dc}$, is divided into two categories: linear and nonlinear. With regard to the nonlinear functions, two subclasses are further classified based on the increasing or decreasing nature of their first-order derivatives. Consequently, we design the two sets of $f(\omega)$ as follows:
	 \begin{equation}
	 	\left\{
	 	\begin{aligned}
	 		f_{1} (\omega) &= \omega,    &  f_{2} (\omega) &= \omega^{3},  &  f_{3} (\omega) &= 1-e^{-3 \omega}, \\
	 		f_{4} (\omega) &= 1-\omega, & f_{5} (\omega)  &= 1-\omega^{3}, & f_{6} (\omega) &= e^{-3 \omega}. 
	 	\end{aligned}
	 \right.
	\end{equation}		
	
	In \tabref{fwb}, we evaluate the NMSE performance of CF-CGN for different variable weight functions. The scenario is set to be LOS \cite{quad} with $\fmin=2$ GHz and $\fmax = 5$ GHz. 
	Compared with 0, which means that we train the network without cycle consistency losses, the extrapolation errors are much more minor under seven other weight functions since $ \mathcal{L}_{cyc} $ couples the paired networks together to explore and utilize the correlation between different bands further. 
	Additionally, it is evident that the overall performance of $ \mathcal{F}_{ic}$ is better than that of $\mathcal{F}_{dc}$ and $\mathcal{F}_{cn}$, and for $f(\omega)\in\mathcal{F}_{ic}$, the estimation error can be as low as 0.0245 ($f_{1}\to f_{2}$) and 0.0256 ($f_{2} \to f_{1}$) when $f(\omega)=\omega$ particularly.
	This result proves that increasing the weight of cycle consistency losses linearly in synchronization with the number of iterations can improve the CF extrapolation accuracy. As a matter of fact, the network will exhibit a large basic loss $\mathcal{L}_{basic}$ in the beginning phases of the iteration, indicating poor extrapolation performance. At this stage, blindly increasing the weight of cycle consistency loss $\mathcal{L}_{cyc}$, like $\mathcal{F}_{dc}$, does not necessarily lead to a significant improvement. However, as the number of iterations increases, the basic loss decreases to a relatively low level. At this time, increasing the weight of cycle consistency loss can further refine the network's performance. Consequently, the weight function needs to be linearly scaled with the number of iterations to achieve optimal network performance.		
	In conclusion, the results presented in \tabref{fwb} underscore the necessity and importance of the variable weight function for cycle consistency losses.
%
	
    \subsection{Evaluation of the Extrapolation Performance}	
    In this section, we adopt two baselines for comparison and evaluate the extrapolation accuracy of CF-CGN  under LOS and NLOS scenarios, where $ f(\omega) $ is set to be $ \omega $. To further validate its generalization ability, we conduct experiments to evaluate the NMSE performance with different numbers of antennas and various frequencies.
    
    \subsubsection{Baselines}
	To compare the performance of CF-CGN, the linear extrapolation-based method \cite{li} and CycleGAN-based algorithm \cite{cyclegan} are adopted as benchmarks. 
	\begin{itemize}
		\item Linear extrapolation-based method \cite{li}: Linear interpolation is one of the most typical traditional extrapolation methods. It usually assumes the same mean angles of arrival (AoAs) and angle spread. Then, it explores the relationship between spatial correlation matrices of different bands by deriving an expression for extrapolation.
		
		
		\item CycleGAN-based method \cite{cyclegan}: Since CycleGAN is one of the most popular approaches for image translation tasks \cite{II}, we adopt it as the other baseline for CF extrapolation. The generator is equipped with eight convolutional layers like the structure in \cite{cyclegan}. Meanwhile, the discriminator comprises five convolutional layers followed by a fully connected layer serving as the output section. The full objective contains three parts: the basic GAN loss \cite{loss}, the cycle consistency loss, and the identity loss \cite{iden}. We conduct the network training based on techniques described in \cite{cyclegan}.
	\end{itemize}
	To ensure fairness in comparison, we utilize the same training and testing datasets for the CycleGAN-based method. Meanwhile, the network parameters, such as learning rates and epochs, are also kept consistent, as shown in \tabref{tb2}.
	
	\subsubsection{Comparison under Different Scenarios}
	In \tabref{different scenarios}, we present the NMSE performance under different communication scenarios: LOS and NLOS.
	In the LOS scenario, our proposed CF-CGN can reduce the estimated error by at least $11.01$ dB compared with the traditional linear interpolation \cite{li}, regardless of whether the extrapolation is performed from $f_{1}$ to $f_{2}$ or from $f_{2}$ to $f_{1}$. Compared with the CycleGAN-based approach \cite{cyclegan}, it also remains a performance advantage of more than $6.4$ dB.
	In the NLOS scenario, the advantage of CF-CGN is even more pronounced. Compared to the traditional linear interpolation and CycleGAN-based approach, when extrapolating from $f_{1}$ to $f_{2}$, our algorithm exhibits advantages of $16.4$ dB and $5.6$ dB, respectively. Conversely, when extrapolating from $f_{2}$ to $f_{1}$, our algorithm displays performance advantages of $17.5$ dB and $8.6$ dB, respectively. These results demonstrate that our proposed CF-CGN can achieve high-precision extrapolation for CF and exhibits good generalization ability in both LOS and NLOS scenarios.	
	
		\newcolumntype{L}{>{\hspace*{-\tabcolsep}}l}
	\newcolumntype{R}{c<{\hspace*{-\tabcolsep}}}
	\definecolor{lightblue}{rgb}{0.93,0.95,1.0}
	\begin{table}[htbp]
		\captionsetup{font=footnotesize}
		\captionsetup{justification=centering}
		\caption{Comparison of the NMSE performance in different scenarios}
		\label{different scenarios}
		\centering
		\ra{1.6}
		\scriptsize
		\begin{tabular}{LcccR}
			\toprule
			\multirow{2}*{Algorithm}         &    \multicolumn{2}{c}{$f_{1}\to f_{2}$}    & \multicolumn{2}{c}{$f_{2}\to f_{1}$} \\	
			\cline{2-5}
			& LOS      &NLOS           & LOS      & NLOS \\
			\midrule
			\rowcolor{lightblue}
			Linear Extrapolation \cite{li}                & 0.3088   &1.4276           & 0.4966  &1.7992             \\
			CycleGAN \cite{cyclegan}                    & 0.1078   &0.1204          & 0.1954   &0.2364             \\		
			\rowcolor{lightblue}
			CF-CGN without refining scheme       &0.0325    &0.0388          &0.0281    &0.0365           \\	
			CF-CGN without $\mathcal{L}_{cyc}$  & 0.0401   &0.0420          &0.0367    &0.0378              \\		
			\rowcolor{lightblue}
			CF-CGN with $\mathcal{L}_{cyc}$ and refining scheme  & \textbf{0.0245}  &\textbf{0.0325}  & \textbf{0.0256}  &\textbf{0.0320}  \\			
			\bottomrule
		\end{tabular}
	\end{table} 

	Furthermore, we also evaluate the influence of cycle consistency losses $\mathcal{L}_{cyc}$ and the refining scheme on the extrapolation results under different scenarios. It is evident that for both LOS and NLOS scenarios,  the design of $\mathcal{L}_{cyc}$  can reduce the extrapolation error by up to $2.2$ dB, which demonstrates the validity and significance of cycle consistency losses. Simultaneously, with the introduction of the refinement mechanism, the extrapolation error can be reduced by, at most, $1.2$ dB, which emphasizes the effectiveness of this refining scheme.
		
	  	\begin{figure*}[htbp]
	  	\centering  
	  	\subfigure[NMSE with $\fmin = 2$ GHz for extrapolation from $f_{1}$ to $f_{2}$.]{
	  		\label{f5}
	  		\includegraphics[width=0.45\linewidth]{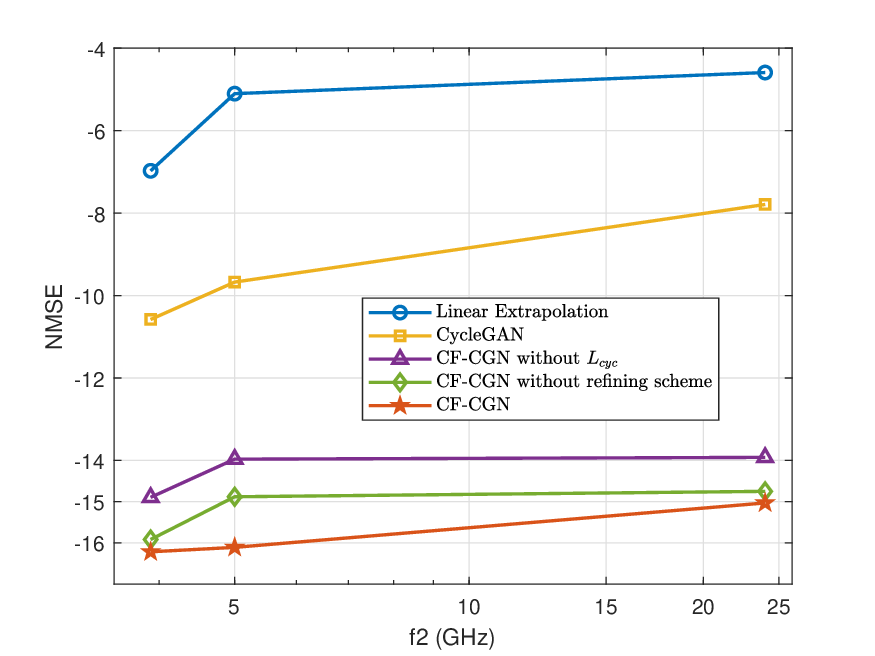}}		
	  		\hspace{0.15 in}
	  	\subfigure[NMSE with $\fmin = 2$ GHz for extrapolation from $f_{2}$ to $f_{1}$.]{
	  		\label{f24}
	  		\includegraphics[width=0.45\linewidth]{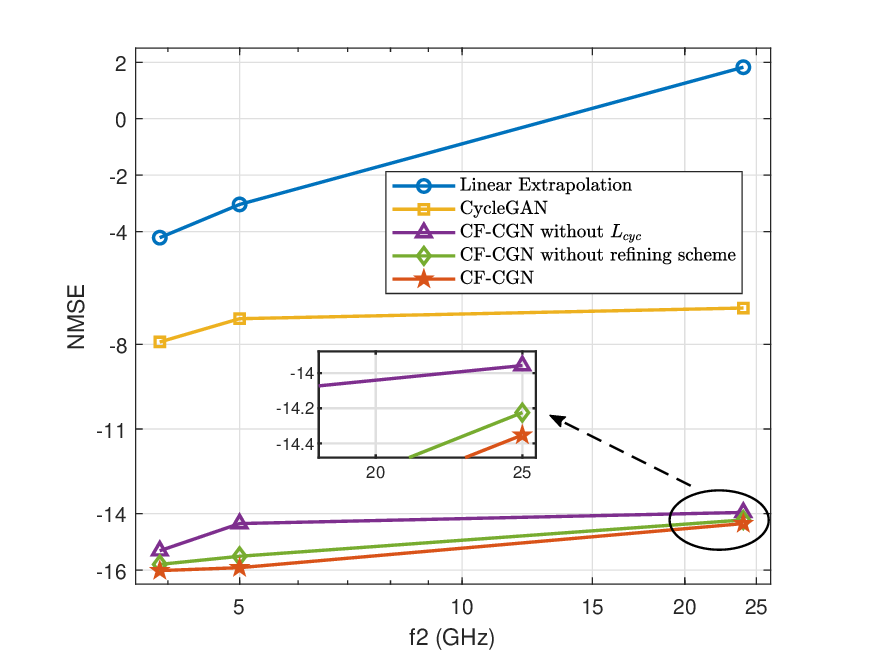}}
	  	\caption{Comparison of the NMSE performance under different frequencies. The communication scenario is set to be LOS with $f_{1}=2$ GHz.}
	  	\label{different frequencies}		
	  \end{figure*}	
	  
	  \begin{figure*}[htbp]
	  	\centering  
	  	\subfigure[NMSE with $M_{1} = 8\times 8$ for extrapolation from $f_{1}$ to $f_{2}$.]{
	  		\label{m10}
	  		\includegraphics[width=0.45\linewidth]{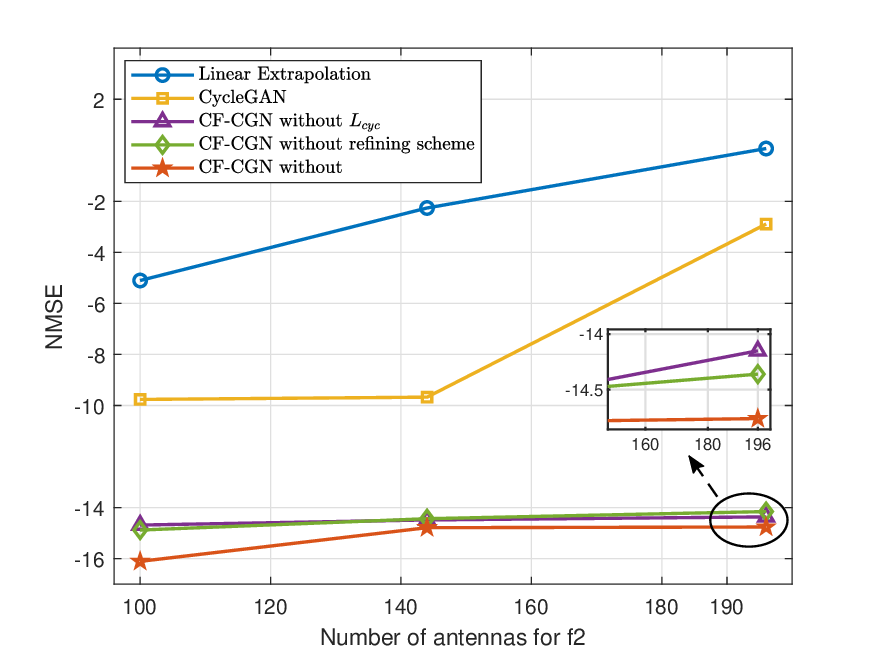}}
	  		\hspace{0.15 in}
	  	\subfigure[NMSE with $M_{1} = 8\times 8$ for extrapolation from $f_{2}$ to $f_{1}$.]{
	  		\label{m12}
	  		\includegraphics[width=0.45\linewidth]{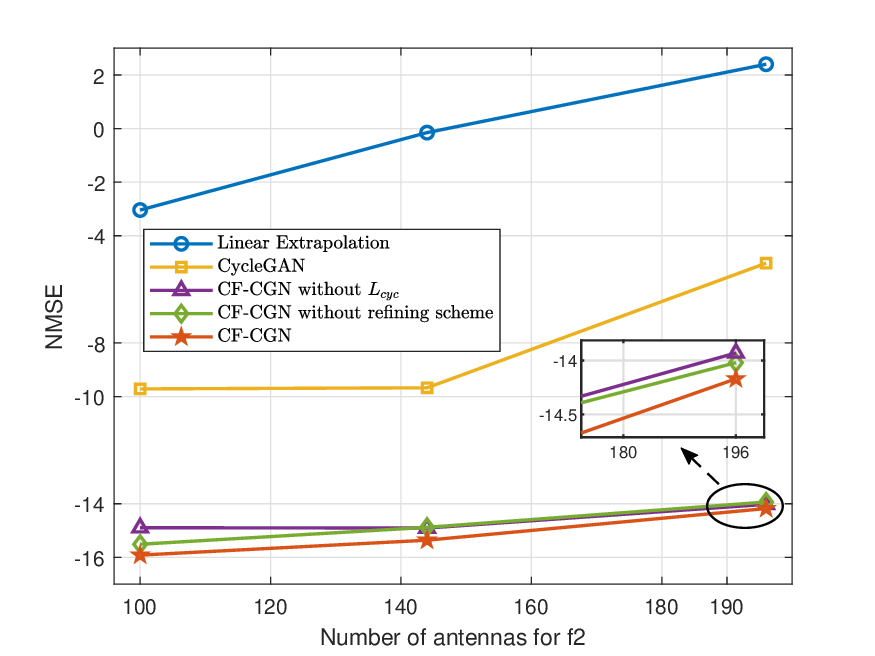}}
	  	\caption{Comparison of the NMSE performance under different number of antennas. The communication scenario is set to be LOS with $f_{1}=2$ GHz and $f_{2}=5$ GHz.}
	  	\label{different number of antennas}		
	  \end{figure*}
	  
	\subsubsection{Comparison under Different Frequencies}
	In \figref{different frequencies}, we show the NMSE performance of CF extrapolation from both $f_{1}$ to $f_{2}$ and $f_{2}$ to $f_{1}$ at different frequencies by our proposed CF-CGN and two baselines. The communication scenario is set to be LOS with $f_{1}=2$ GHz. For the design of $ f_{2} $, we choose three typical frequencies from sub-6GHz, unlicensed spectra in WiFi, and mmW bands, respectively, including $3.6$ GHz, $5$ GHz, and $24$ GHz. The extrapolation error increases as $ f_{2} $ changes from $3.6$ to $24$ GHz. In practice, a more significant gap between $ f_{1} $ and $ f_{2} $ will lead to more differences in their electromagnetic propagation characteristics, weakening the correlation between the statistical CSI and decreasing the extrapolation accuracy. However, regardless of how $ f_{2} $ changes, our proposed CF-CGN consistently performs better than the other baselines, demonstrating its applicability across different frequencies.
	
	\subsubsection{Comparison under Different Numbers of Antennas}
	In \figref{different number of antennas}, we provide the NMSE performance under different numbers of antennas for $f_{2}$. The communication scenario is set to be LOS with $f_{1}=2$ GHz and $f_{2}=5$ GHz. The antenna array of $f_{1}$ is set to be $M_{1}=8\times 8$ while that of $f_{2}$ is $M_{2}=10\times 10$,  $12\times 12$, and $15\times 15$. As depicted in \figref{different number of antennas}, the extrapolation accuracy gradually decreases with the increase of $M_{2}$. \cite{pp} has pointed out that in multi-band massive MIMO systems, the percentage of overlapping beams decreases as the number of antennas at $f_{2}$ increases, weakening the correlation between two bands and thus deteriorating the extrapolation performance. Nevertheless, our proposed CF-CGN can still maintain the error at approximately $-14$ dB, outperforming other benchmark algorithms. 
	
				\begin{figure*}[htbp]
		\centering  
		\subfigure[Comparison of the sum rate performance in LOS scenarios with $f_{2}=5$ GHz and $M_{2} = 10\times 10$.]{
			\label{los}
			\includegraphics[width=0.45\linewidth]{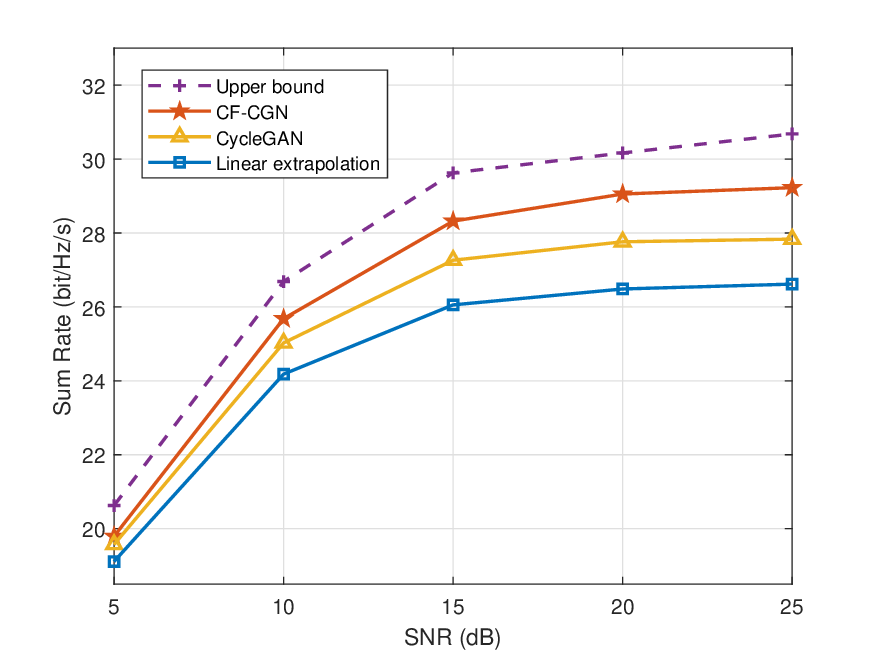}}
		\hspace{0.15 in}
		\subfigure[Comparison of the sum rate performance in NLOS scenarios with $f_{2}=5$ GHz and $M_{2} = 10\times 10 $.]{
			\label{nlos}
			\includegraphics[width=0.45\linewidth]{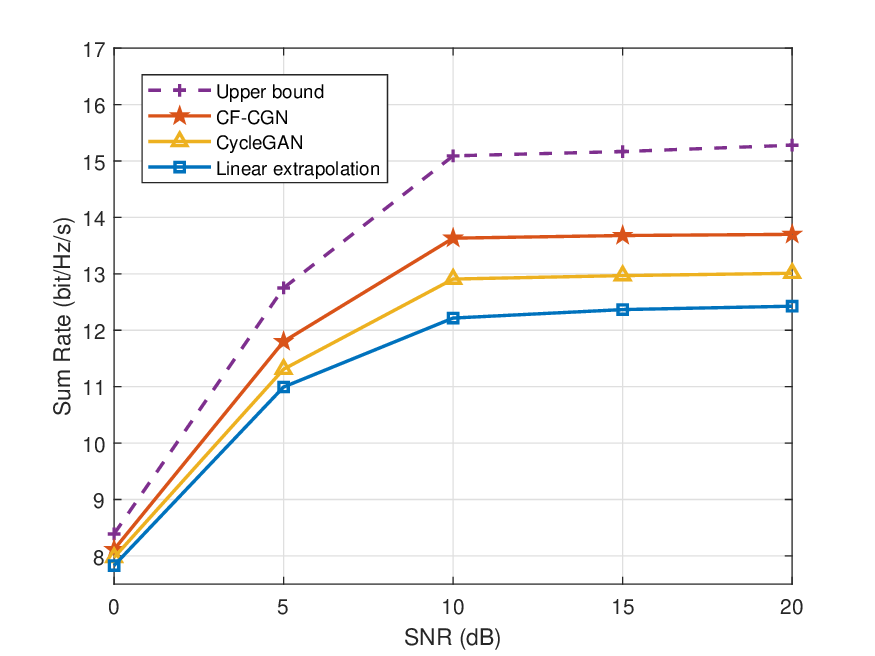}}
		
		\subfigure[Comparison of the sum rate performance in LOS scenarios with $f_{2}=5$ GHz and $M_{2} = 12\times 12$.]{
			\label{12}
			\includegraphics[width=0.45\linewidth]{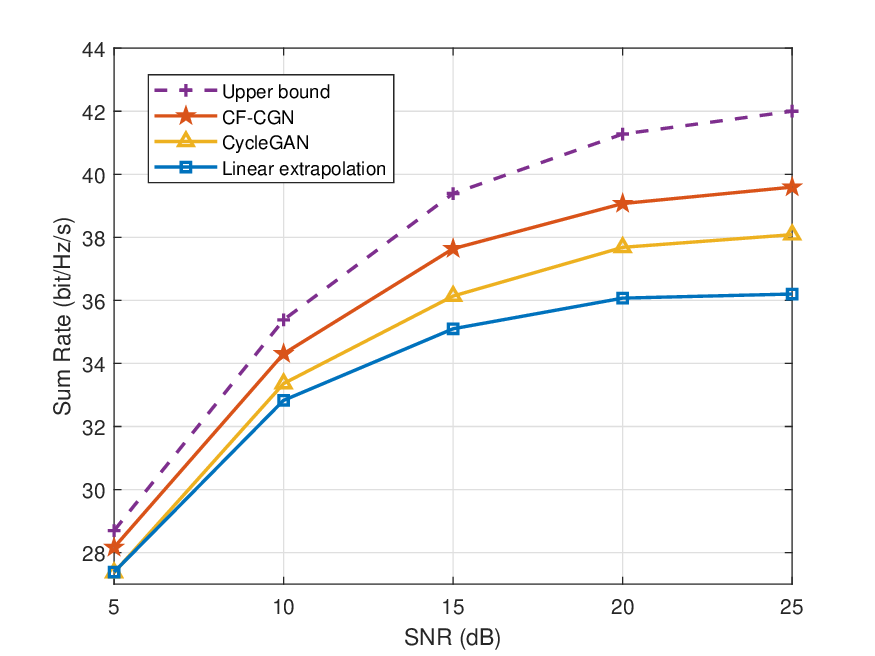}}
		\hspace{0.15 in}
		\subfigure[Comparison of the sum rate performance in LOS scenarios with $f_{2}=24$ GHz and $M_{2} = 10\times 10 $.]{
			\label{24}
			\includegraphics[width=0.45\linewidth]{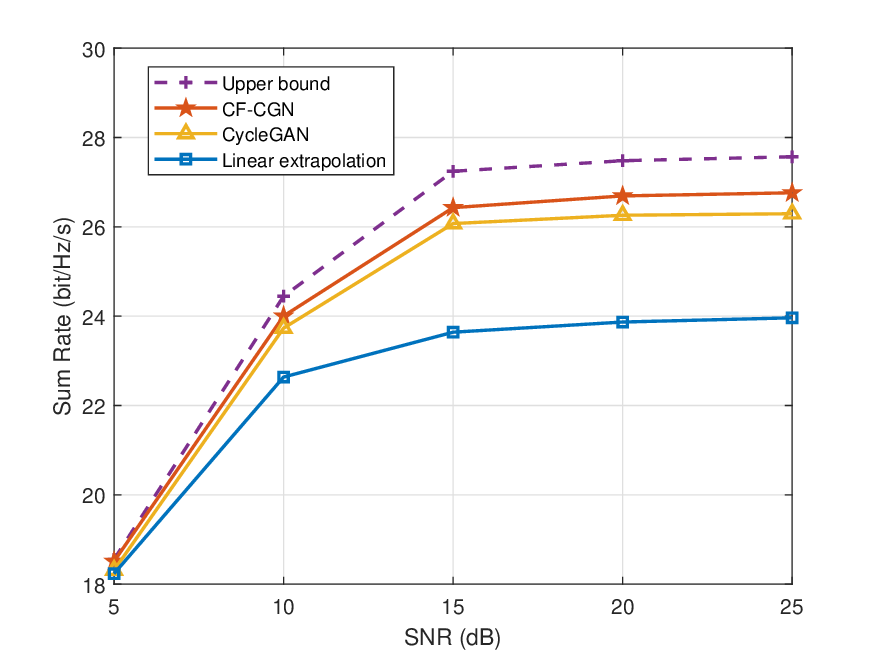}}                            
		\caption{Comparison of the sum rate performance with robust precoding for multi-band massive MIMO transmission.}
		\label{rrrr}        
	\end{figure*}
	
		\subsection{Multi-band Massive MIMO Transmission}			
		To further validate the multi-band massive MIMO transmission performance based on the extrapolated CF, we calculate the sum rate of the downlink transmission, an important indicator to measure communication quality. Notably, robust transmission under nonideal conditions can be achieved by precoding matrices generated from the discrete PAS \cite{lby}. Therefore, we construct the precoding matrix $\mathbf{P}_{l}$ based on the extrapolated PAS to validate the performance of CF-assisted multi-band massive MIMO transmission. For the $l$-th UT that is classified into the $(m, n)$-th grid, the transmission rate is given by \cite{tjk}
		\begin{equation}
			R_{l} = \mean \{ \log_{2} \det({{\bf{V}}_{l}} + \mathbf{H}_{l} \mathbf{P}_{l} \mathbf{P}_{l}^{H} \mathbf{H}_{l}^{H}) - \log_{2}\det{{\bf{V}}_{l}} \},
			\label{rl}
		\end{equation}
		where $\mathbf{H}_{l}$ denotes the channel matrix between BS and the $l$-th UT, and $\mathbf{P}_{l}$ is the precoding matrix. ${{\bf{V}}_{l}}$ here represents the covariance matrix of channel interference that encapsulates both the additive noise and the inter-user interference, which is given by
		\begin{equation}
			{{\bf{V}}_{l}} = \sigma_{l}^{2} + \sum_{l'\ne l}^{L_{\rm total}} \mean \{ \mathbf{H}_{l'} \mathbf{P}_{l} \mathbf{P}_{l}^{H} \mathbf{H}_{l'}^{H} \},
		\end{equation}
		where $\sigma_{l}^{2}$ is the variance of a complex Gaussian noise and $L_{\rm total}$ represents the number of users.	
		Based on (\ref{rl}), we can give the sum rate by
		\begin{equation}
			R = \sum_{l=1}^{L_{\rm total}} R_{l}.
		\end{equation}
		Simulations are conducted under four distinct scenarios, encompassing various numbers of antennas, different frequencies, and diverse propagation environments (LOS and NLOS). Specifically, we first obtain the statistical CSI of $f_{1}$ through uplink detection and construct the CF accordingly. Then, we extrapolate CF by linear extrapolation method, CycleGAN-based method, and our proposed CF-CGN to acquire the PAS at $f_{2}$. Subsequently, the precoding matrix is designed based on the robust approach in \cite{lby}, thus enabling the multi-band cooperative transmission. 
		
	\figref{rrrr} presents the sum rate performance with robust precoding for multi-band massive MIMO transmission. It is evident that regardless of the communication scenarios, channel fingerprints extrapolated by our proposed CF-CGN can always assist the multi-band massive MIMO system in achieving the highest sum rate.
	Besides, in NLOS scenarios shown in \figref{nlos}, the distribution of scatterers becomes more complex, so the accuracy of the extrapolated CF decreases as displayed in \tabref{different scenarios}, which restricts the further improvement of the sum rate.
	Variations in the number of antennas also influence the CF-assisted multi-band massive MIMO transmission. When $M_{2}=144$, the increase of the channel dimension elevates the difficulty of CF extrapolation, so the sum rate assisted by CF-CGN-based CF can only reach $90$\% of that with the idea CSI. However, it still outperforms the CycleGAN-based method and the linear extrapolation approach.
	Furthermore, as $f_{2}$ transitions into the mmW band, the discrepancies in electromagnetic propagation characteristics between two frequencies progressively expand. Consequently, shown in \figref{24}, the performance of the traditional linear extrapolation method deteriorates sharply, resulting in a sum rate that reaches only $85$\% of the ideal value. In contrast, the CF-CGN-based method still maintains a performance level of $96$\%. It is also worth noting that for mmWave transmission, the LOS link will dominate among the multipath components. Although the overall extrapolation error of CF-CGN is lower than that of CycleGAN, their estimated powers in the LOS direction are comparable, resulting in a relatively small difference in the transmission rate after robust precoding. 
	To conclude, \figref{rrrr} demonstrates that the CF-CGN-based CF can effectively assist the multi-band massive MIMO transmission and achieve a higher sum rate than benchmarks.

	\subsection{Complexity Comparison}
	Based on the network structure in \figref{net} and its parameters in \tabref{tb2}, we compare the time complexity and storage complexity for our proposed CF-CGN in the multi-band CF extrapolation task. 
	Since the time complexity of CF-CGN primarily depends on the computations of convolutional layers, we express it as $ \sum_{\ell=1}^{L} \mathcal{O}_{\ell}(H_{\ell}W_{\ell}C_{\ell}C_{\ell-1}K^{2}_{\ell}) $, where $ H_{\ell}, W_{\ell}$, and $ C_{\ell} $ are the heights, widths, and channels of the $ \ell $-th layer's input, respectively, and $ K_{\ell} $ is size of the corresponding convolution kernel. 
	The space complexity is mainly reflected in the number of network parameters related to the data channels and the convolution kernel size. Therefore, we can express it as $ \sum_{\ell=1}^{L} \mathcal{O}_{\ell}(C_{\ell}C_{\ell-1}K^{2}_{\ell}) $.
	
	\newcolumntype{L}{>{\hspace*{-\tabcolsep}}l}
	\newcolumntype{R}{c<{\hspace*{-\tabcolsep}}}
	\definecolor{lightblue}{rgb}{0.93,0.95,1.0}
	\begin{table}[htbp]
		\captionsetup{font=footnotesize}
		\captionsetup{justification=centering}
		\caption{Analysis of the model complexity}\label{com}
		\centering
		\ra{1.5}
		\scriptsize
		\begin{tabular}{LccR}
			\toprule
			                & Parameters  \ \ \ \ \ \ & FLOPs \ \ \ \ \ \ \ & Memories \\
			\midrule
			\rowcolor{lightblue}
			CF-CGN   & 126.47 M \ \ \ \ \ \   & 5.71 G  \ \ \ \ \ \ \ & 29.65 MB \\
			CycleGAN \ \ \ \ \ \ \ & 184.64 M \ \ \ \ \ \   & 15.57 G \ \ \ \ \ \ \ & 48.25 MB \\
			\bottomrule
		\end{tabular}
	\end{table}

     \tabref{com} gives the specific numerical results of parameters, floating point operations (FLOPs), and memories. Compared with the CycleGAN-based method, our proposed CF-CGN has less trainable parameters, FLOPs, and required memory space, making the model more portable and efficient. It is worth noting that the complexity of CF-CGN stems from the introduction of cycle consistency loss, which can trade additional computational costs in exchange for the ability of bidirectional prediction and reduce estimation errors. Since the ML model is typically deployed on high-performance devices such as BSs, the computational requirements of CF-CGN are entirely acceptable.

	\section{Conclusion}\label{sec_conclusion} 
	In this paper, we proposed CF-CGN to extrapolate CF for multi-band massive MIMO systems based on the cycle-consistent generative networks, where licensed and unlicensed spectra cooperate to provide ubiquitous connectivity.
	Firstly, we established the multi-band channel model and revealed a relationship between the spatial covariance matrix and PAS. By viewing CF as a multichannel image, we investigated the CF model and transformed the extrapolation problem into an image translation task, which aims to convert CF from one frequency to another by exploring the shared characteristics of statistical CSI in the beam domain. Then, we designed a pair of generative networks and coupled them by variable-weight cycle consistency losses to fit the reciprocal relationship of different bands. Then, a joint training strategy was developed accordingly, supporting synchronous optimization of all trainable parameters. During the inference process, we also introduced a refinement mechanism based on the pixel-to-grid conversion to improve the extrapolation accuracy.
	Numerical results illustrated the effectiveness and generalization ability of CF-CGN under different communication scenarios. We also demonstrated that the sum rate performance for CF-CGN-assisted multi-band massive MIMO transmission is very close to that with perfect CSI.

	\bibliographystyle{IEEEtran}
	\bibliography{EE_AI}
	
\end{document}